\documentclass{aa}  

\usepackage{graphicx}
\usepackage[switch]{lineno} 
\usepackage{txfonts}
\usepackage{arydshln}
\usepackage{pdflscape}
\usepackage{booktabs}
\usepackage{changes}
\usepackage[flushleft]{threeparttable}
\usepackage[colorlinks,allcolors=blue]{hyperref}

\newcommand{\RomanNumeralCaps}[1]
    {\MakeUppercase{\romannumeral #1}}

\makeatletter
\@namedef{Changes@AuthorColor}{orange} 
\colorlet{Changes@Color}{orange}
\makeatother

\begin{document} 

   \title{A background-estimation technique for the detection of extended gamma-ray structures with IACTs}

   \author{ T. Wach \inst{1} \and
            A. Mitchell \inst{1} \and
            L. Mohrmann \inst{2}}

   \institute{Friedrich-Alexander-Universit\"at Erlangen-N\"urnberg, Erlangen Centre for Astroparticle Physics, Nikolaus-Fiebiger-Str. 2, 91058 Erlangen, Germany\\
              \email{tina.wach@fau.de}
         \and
             Max-Planck-Institut f\"ur Kernphysik, Saupfercheckweg 1, 69117 Heidelberg, Germany\\}


 
  \abstract
   {Estimation of the amount of cosmic-ray induced background events is a challenging task for Imaging Atmospheric Cherenkov Telescopes (IACTs). 
   Most approaches rely on a model of the background signal derived from archival observations, which is then normalised to the region of interest (ROI) and respective observation conditions using emission-free regions in the observation.
   This is, however, disadvantageous for the analysis of large, extended $\gamma$-ray structures, where no sufficient source free region can be found. 
   }
   {We aim to address this issue by estimating the normalisation of a 3-dimensional background model template from separate, matched observations of emission-free sky regions. As a result, the need for a emission-free region in the field of view of the observation becomes unnecessary.}
   {For this purpose, we implement an algorithm to identify observation pairs with as close as possible observation conditions. The open-source analysis package \texttt{Gammapy} is utilized for estimating the background rate, facilitating seamless adaptation of the framework to many $\gamma$-ray detection facilities. Public data from the High Energy Stereoscopic System (H.E.S.S.) is employed to validate this methodology.}
   {The analysis demonstrates that employing a background rate estimated through this run-matching approach yields results consistent with those obtained using the standard application of the background model template. Furthermore, the compatibility of the source parameters obtained through this approach with previous publications and an analysis employing the background model template approach is confirmed, along with an estimation of the statistical and systematic uncertainties introduced by this method.}
   {}


   \maketitle
%

\section{Introduction}
Imaging Atmospheric Cherenkov Telescopes (IACTs) have considerably expanded our knowledge of the very high energy $\gamma$-ray sky. With the advent of the upcoming Cherenkov Telescope Array \citep[CTA]{cta}, IACTs will remain a crucial tool for $\gamma$-ray astronomy for the foreseeable future. Despite this, recent results from Water Cherenkov Detectors (WCDs), such as the High-Altitude Water Cherenkov Array \citep[HAWC]{hawc} and Large High Altitude Air Shower Observatory \citep[LHAASO]{lhaaso}, have highlighted a limitation of IACTs. Whilst their superior angular resolution enables IACTs to distinguish between $\gamma$-ray sources situated in close proximity to one another and measure the extension of those sources with high precision, identifying large extended structures of $\gamma$-ray emission remains challenging. 
\\
This is primarily due to the effect the comparatively small field of view (FoV) of IACTs has on the detection of extended $\gamma$-ray sources \citep{hess-hawc}. While WCDs are survey instruments that continuously monitor the overhead sky, IACTs need to be pointed towards a target source and can only observe the sky in a region a few degrees across. This affects the estimation of background rates in the case of studying extended source regions that cover a large part of the FoV. The background consists of non-$\gamma$-ray induced extensive air showers, predominantly due to hadrons interacting with the Earth's atmosphere, albeit at lower energies cosmic-ray electrons also play a significant role. By implementing selection criteria on the reconstructed shower parameters \citep{gamma-hadron}, this background can be significantly reduced, but not fully removed. The residual background rate is then often estimated from source-free regions in the FoV of the observation \citep{on-off-method}. \footnote{Another approach, initially proposed by \citet{template}, is to estimate the background rate from a separate set of events that are more cosmic ray-like than the selected gamma-ray candidates. This approach requires detailed knowledge about the acceptance of the system to these cosmic ray-like events, however, and will not be discussed further in this work.}
\\
However, it is not always possible to find a sufficiently large source-free region, especially in the case of large, extended sources filling a significant fraction of the FoV, which can lead to the subtraction of significant emission in the whole FoV \citep{geminga, J1813, hess-hawc}. The choice of a region that is not free of gamma-ray emission can then lead to absorption of these structures in the background estimation. 
To circumvent this problem, a spectromorphological background model template, constructed from archival data, enabeling a three-dimensional (3D) likelihood analysis, can be used (for more information see \citet{bkg_template}). 
Due to the large amount of observations, acquired under similar observation conditions, used for the creation of the background model template, this approach is very stable and suffers only little from statistical fluctuations in the background estimate. Previous studies have shown that the background estimation employing a background model template facilitates the detection of large, extended structures with IACTs \citep{hess-hawc}. However, the background model template approach still requires a re-normalisation for each observation run, to account for differences in atmospheric conditions or slight hardware degradation, again requiringat least a small emission-free region in the FoV.
\\
This problem was highlighted in a recent study of emission around the Geminga pulsar (PSR~J0633$+$1746) with the High Energy Stereoscopic System (H.E.S.S.) \citep{geminga}. The study revealed that while it is possible to detect the extended emission around the pulsar, an absolute measurement of its properties was not possible due to the lack of emission-free regions in the FoV \citep{geminga}. Instead, only a relative measurement was feasible, where the background level was estimated in regions of fainter $\gamma$-ray emission.
\\
In regions where no emission-free region can be identified, a so-called ``ON/OFF'' approach can be used, whereby every observation of the targeted source (ON run) is matched to an observation conducted in a part of the sky that is predominantly source-free (OFF run). Such an analysis is for example described in \citet{on-off-pks}
\\
This approach is a standard technique for background rejection in IACTs \citep{on-off-method}, which was already employed by the Whipple observatory \citep{whipple}. Though this standard approach can be very powerful for the observation of extended sources, because no assumptions regarding the background acceptance across the FoV needs to be made, it also has major disadvantages. In the classical approach, an empty sky region needs to be observed for a comparable duration and under a similar zenith angle (with an allowed Zenith angle deviation between ON and OFF run of typically $30’$) \citep{on-off-method, on-off-pks}. Additionally, the best description can be achieved if the OFF run is recorded shortly after the ON run, in order to avoid the influence of effects like degradation of the system or changes in the atmosphere. This means that at most half of the observation time can be spent on the actual target of the study. Additionally, a background estimate based on only one observation run is necessarily afflicted with relatively large statistical uncertainties.
\\
In this work, we alleviate the disadvantages of both background estimation techniques by combining the classical ON/OFF method with the 3-dimensional background model template constructed from archival data. For this purpose, the normalisation of the background model template for each run pair is determined from the OFF run and this normalisation is then used for the ON run. Employing this background model template differs from the classical approach, since the background rate is not only estimated from one OFF run, but many. This enables us to decrease the dependence of the background rate on the particular OFF run selected and thus to lower the uncertainty of the background estimate. 
\\
The new method is developed using the data structure of the High Energy Stereoscopic System (H.E.S.S.), an array of five IACTs located in the Khomas Highland in Namibia \citep{instrument_systematics, hess-recent}. The original telescope array consisted of four telescopes with $12\,$m-diameter mirrors (CT1-4) and was commissioned between 2000 and 2003. In 2012 a fifth telescope (CT5), with a mirror diameter of $28\,$m was added in the centre of the $120\,$m square spanned by CT1-4, lowering the energy threshold of the array considerably \citep{crab}. The method was then validated using data from the first H.E.S.S. public data release \citep{data-release}. Since this method only requires common python packages, as well as \texttt{Gammapy} \citep{gammapy_zenodo, gammapy}, it is in accordance with the goal of the Very-high-energy Open Data Format Initiative (VODF; \citealt{VODF}), to achieve a common, software-independent data format and develop open source analysis software \citep{GADF}. This also means that the method can easily be adapted for other telescope arrays. 
\\
Henceforth, we will refer to the template constructed in \citet{bkg_template} as `background model template', the standard method employed by the Whipple Observatory as `classical ON/OFF method' \citep{whipple}, and the method developed in this work as `run-matching approach'. With this publication, we will release the scripts necessary to perform the background matching, after a set of OFF runs has been selected, as well as the results for all spectral and spatial fits on the public-dataset release. 
\\
For this study, only data acquired by the four smaller telescopes of the H.E.S.S. array is analysed. For all observations used, the data reduction was performed using HAP, the H.E.S.S. analysis package described in \citealt{instrument_systematics}, and reconstructed using the Image Pixel-wise fit for Atmospheric Cherenkov Telescope algorithm \texttt{ImPACT} \citep{impact}.

\section{Run-matching}
\label{matching}
For the classical ON/OFF method, an observation of an empty region with similar properties to that of the target region needs to be conducted. This requirement is necessary to achieve a comparable system acceptance (relative rate of events passing selection cuts). While the use of a background model template allows for relaxed run-matching criteria with respect to the classical ON-OFF matching, allocating a comparable OFF run is nevertheless a critical aspect of the background estimation. 

\begin{table}[ht]
\caption{The binning used for the background model template.} 
\label{tab:zenith_bins} 
\centering                          
\begin{tabular}{c | c }       
\hline\hline  
\noalign{\smallskip}
$\Theta_z\, [^\circ]$ &$N_\text{obs}$  \\    
\noalign{\smallskip}
\hline 
\noalign{\smallskip}
   $ 0 - 10$  & $99$ \\
   $10 - 20$  & $392$ \\
   $20 - 30$  & $650$ \\
   $30 - 40$  & $444$ \\
   $40 - 45$  & $300$ \\
   $45 - 50$  & $306$ \\
   $50 - 55$  & $150$ \\
   $55 - 60$  & $61$ \\
\noalign{\smallskip}
\hline   
\hline 
\end{tabular}
\tablefoot{The number of observations $\text{N}_\text{obs}$ per zenith angle $\Theta_z$ bin, taken from \cite{bkg_template}.}
\end{table}

Previous works employing the classical ON/OFF method have shown that a variety of parameters influence the background rate and need to be considered in the matching process \citep{matching_params1, matching_params2, geminga, thesis_veh}. In this work, an OFF run is only considered if its pointing position is at a galactic latitude of $|b| \geq 10^\circ$, in order to avoid regions including many known $\gamma$-ray sources, as well as diffuse $\gamma$-ray emission. Furthermore, we only consider observations taken under good atmospheric conditions, and a good system response, and require that the same telescopes of the array participate in both the ON and OFF run. This quality selection was performed following the recommendations in \citet{instrument_systematics}.

\subsection{Matching parameters}
The parameters found to have the largest influence are the zenith angle (the angle between the pointing direction of the telescopes and zenith), changes in hardware configuration and atmospheric conditions. The level of night sky background (NSB) light is also used for matching, due to its influence on the performance of the telescopes \citep{magic_nsb, veritas_nsb}. Since small changes of the atmosphere and degradation of the system can be absorbed by the fit of the background model template (see section \ref{background-fit}), the run pairs do not need to be acquired in the same night, but can be up to a few years apart. It is however important to match runs using the same hardware configuration. Therefore we take optical phases into account. These are periods of stable optical efficiency, between abrupt changes due to, for example, cleaning of the Winston cones (light guides attached to the camera pixels) or mirror re-coating. Typically these phases span at least one year for the H.E.S.S. telescopes, therefore enabling the choice of many possible OFF runs for one ON run. The optical phases used for this work can be found in Table \ref{tab:zenith_correction}.
\\
\begin{table*}[ht]
\caption{The matching parameters used for the OFF run estimation.} 
\label{tab:params_deviations} 
\centering                          
\begin{tabular}{c  | c  c c c}       
\hline\hline  
\noalign{\smallskip}
Matching parameter & Valid parameter range & \multicolumn{3}{c}{Correlation $d_{corr}$} \\    
\noalign{\smallskip}
\hline 
\noalign{\smallskip}
                    &                  & HESS Phase 1 & HESS Phase 2 & HESS Phase 1U \\
    Zenith Angle $\Theta_z$ & within zenith bins & 0.36 & 0.32 & 0.45\\
    Trigger Rate $r$ & $\Delta\, r < 25\%$    & 0.42 & 0.66 & 0.46  \\
    Duration $t$ & $\Delta\,t < 7\%$          & 0.64 & 0.68 & 0.71 \\
    Transparency Coefficient $\tau$ & $\Delta\,\tau < 6\%$ & 0.43 & 0.66 & 0.55 \\
    \noalign{\smallskip}
    \hline
    \noalign{\smallskip}
    Radiometer Temperature $R_T$ & $\Delta\,R_T < 50\%$ & 0.23 & 0.51 & 0.48 \\
    Muon Efficiency $\varepsilon_\mu$ & $\Delta\,\varepsilon_\mu < 11\%$ & 0.42 & 0.37 & 0.16 \\
    NSB & $\Delta\, \text{NSB} < 80\%$        & 0.32 & 0.25 & 0.24 \\
\noalign{\smallskip}
\hline   
\hline 
\end{tabular}
\tablefoot{Additionally given are the respective allowed deviations of all parameters and their distance correlation \citep{distance_cov} with the background rate.}
\end{table*}
\\
For the construction of the background model template the zenith angles of the OFF runs were grouped into bins with a size dependent on the available statistics, because of the strong influence of the zenith angle on the background rate \citep{zenith_bkg_proof}. The zenith angle bins from \citet{bkg_template} were adopted as a validity range of the zenith angle deviation for this work, to ensure compatibility with the background model template (see Table \ref{tab:zenith_bins}). In order to have a comparable background model template between ON and OFF run, the azimuth angle bins from \citet{bkg_template} where also adopted for this study. 
\\
Another important matching parameter is the so-called muon efficiency $\varepsilon_\mu$. This quantity specifies how many photo-electrons are detected per incident photon, and is, therefore, a measure of the optical performance of the telescopes \citep{muon1, muon2}. Its estimation through measurements on muon events becomes possible since the geometry of the muon images can be used to calculate the expected intensity, which can then be compared to the measured intensity. The second matching parameter introduced to account for atmospheric changes is the

To evaluate the differences in atmospheric conditions between two runs, two quantities were investigated. The first parameter is the Cherenkov transparency coefficient \citep{transp_coeff_calc}. It describes the transparency of the atmosphere and can be calculated via:
\begin{equation}
\centering
    \tau = \frac{1}{N \cdot k_N} \sum_i t_i = \frac{1}{N \cdot k_N} \sum_i \frac{R_i^\frac{1}{1.7-\Delta}}{\mu_i \cdot g_i}
\end{equation}
with the number of participating telescopes $N$, the average amplification gain of the photosensors $g_i$, and the trigger rate $R_i$ and muon efficiency $\mu_i$ of observation $i$. The term $\Delta$ allows for higher order corrections and $k_N$ is a scaling factor \citep{transparency}. The second matching parameter is the effective sky temperature in the FoV of the individual telescopes, measured with infrared radiometers \citep{radiometer}. Following this quantity will be referred to as radiometer temperature. 
\\
Additionally, we require the ON and OFF runs to have a comparable dead-time-corrected observation time.
\\
In this study, the influence of the respective matching parameters on the background estimate has been estimated by calculating the distance correlation \citep{distance_cov} between the matching parameters and the number of background events estimated in an OFF region using the background model template. In contrast to the Pearson correlation coefficient, the distance correlation allows the correlation between two parameters to be measured independently of the nature of their correlation.
\\
Since the influence of the matching parameters can vary following major changes in the hardware configuration of the telescope array, the correlation coefficients have been calculated for each of the three hardware phases of the H.E.S.S. telescopes. HESS Phase 1 represents the data taken from the commissioning of the first four telescopes until the addition of the fifth telescope in 2012 \citep{instrument_systematics}. The data taken with the five telescope array is called HESS Phase 2 \citep{hess_phase2}. The last set, called HESS Phase 1U, includes all data taken after the camera upgrade of the four small telescopes in 2017 \citep{hess1u}.
\\
For a correlation coefficient of $d_{corr} \geq 0.15$, the influence of the matching parameter on the background rate was regarded as significant. A summary of all significant matching parameters, as well as allowed deviations and correlation coefficients for this work can be found in Table \ref{tab:params_deviations}. 
\\
These correlation coefficients were estimated using the number of background counts over the whole energy range. We additionally computed correlation coefficients using only low-energy events and find that there is no significant differences to the coefficients presented in Table \ref{tab:params_deviations}.

\subsection{Fractional run difference}
The deviations of the matching parameters are then used to quantify the difference between an ON and OFF run, by calculating the fractional run difference $f$:
\begin{equation}
    f = \sum_j d_{corr,j} \cdot \frac{x_\text{on}^j - x_\text{off}^j}{x_\text{on}^j}
\end{equation}
with $j$, one of the matching parameters and $d_{corr,j}$ the distance correlation for the respective matching parameter (see Table \ref{tab:params_deviations}). The observation with the smallest fractional run difference is chosen as the OFF run for the corresponding ON run. 
\\
If no OFF run fulfilling all matching criteria can be found, only the parameters with the largest influence (top half of Table \ref{tab:params_deviations}) were used for the matching. 
In this case, the fractional run difference computed with all matching parameters can be large, and the sub-optimal matching is then taken into account by increased systematic errors (as will be described in Section \ref{sys_errors}).


\section{Background estimation}\label{background-fit}
\subsection{General Method}
After matching every ON run with an OFF run, the background model template was normalised to the OFF run. During this process, the background rate $R_{BG}$ underwent correction for minor discrepancies stemming from varying observation conditions, employing $R^*_{BG} = \Phi \cdot R_{BG} \cdot (E/E_0)^{-\delta}$, where $E_0 = 1\,$TeV is the reference energy and the spectral tilt $\delta$ and background normalisation $\Phi$ are determined through a 3D likelihood fit of the background model template to a emission-free region in the OFF run \citep{bkg_template}. 
\\
For every ON and OFF run, the energy at which the deviation between true and reconstructed energy, the energy bias, reaches 10\%, was estimated. This energy was then set as a safe energy threshold and the data at lower energies discarded. The maximally allowed offset between the reconstructed event direction and the pointing position of the camera was $2.0^\circ$.
\\
\\
\begin{figure}
\includegraphics[width=9cm]{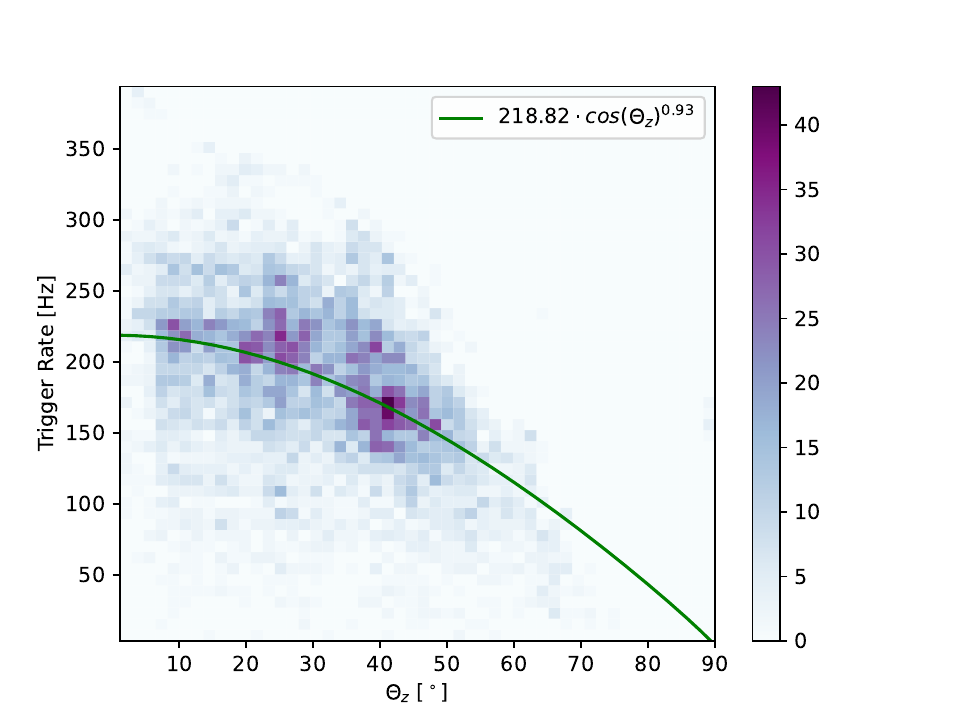}
\caption{The dependence of the array trigger rate of the four small H.E.S.S. telescopes on the zenith angle of the measurement for the first optical phase. A fit of equation \eqref{eqn:zenith_correction} to the data yields the parameters used for the background correction.}
\label{fig:zenith_factor}
\end{figure}
\texttt{Gammapy version 1.2} \citep{gammapy_zenodo, gammapy} was used to create a dataset with a square region geometry of $4^\circ \times 4^\circ$ centred around the pointing position of the OFF run. This \texttt{MapDataset} combines information such as a counts map (the observed number of events passing selection cuts in each bin), expected background map, and exposure. The size of the geometry was chosen such that all of the data passing the thresholds was included in the analysis. For a correct description of the amount of cosmic ray background in the observation, it is important to ensure that no $\gamma$-ray source is present in the region to which the parameters of the background model template are fitted. 
\\
Therefore, the regions containing previously identified extragalactic $\gamma$-ray sources in the OFF run were excluded from the fit of the background model template. Additionally, we excluded a circular region with a radius of $0.5^\circ$ around the observation target of the respective OFF run, to minimise the amount of $\gamma$-ray emission included due to sub-threshold $\gamma$-ray sources. The background model template was then fit to the OFF run. Thereafter, only the adjusted background model template was used.
\\
Additionally to minimising the deviations between ON and OFF run by comparing the fractional run deviation and identifying the best-matching OFF run, deviations between the observations can be accounted for by applying correction factors. 
\\
One correction applied accounts for the deviation in duration (observation time during which the telescope system can trigger on a signal) between ON and OFF run:
\begin{equation}
    b_{i,ON} = \frac{b_{i,OFF} \cdot t_{ON}}{t_{OFF}} 
\end{equation}
with $b_i$ the total number of background events per spatial pixel and $t$ the total duration of the observation. 
\\
The second correction applied accounts for the deviation in zenith angle of the observation. Because of the matching within the zenith angle bins used in the computation of the background model template (see Table \ref{tab:zenith_bins}), some run pairs can have a deviation of up to $5^\circ$ in zenith angle $\Theta_z$. The correction factor to account for this deviation is calculated using:
\begin{equation}
\label{eqn:zenith_correction}
    b_{ON} = b_{OFF} \cdot p_1 \cos{(\Theta_z)}^{p_2} \,.
\end{equation}
This correction follows the relation between the cosmic ray rate, and therefore also the trigger rate, which is dominated by hadronic background events, and the zenith angle of the measurement established in \citet{zenith_bkg_relation}. The parameters $p_1$ and $p_2$ were estimated for every optical phase, by fitting equation \eqref{eqn:zenith_correction} to the trigger rate of the H.E.S.S. array for all observations performed in the respective optical phase. An example of such a fit to the trigger rates can be seen in Figure \ref{fig:zenith_factor}. The fit parameters for all optical phases, as well as more information about their computation, can be found in Appendix \ref{zenith_correction_sec}.
\\
In the next step, another \texttt{MapDataset} was created and filled with the reconstructed events observed in the ON run. The background model template was then assigned to this dataset, and the spectral tilt $\delta$ and background normalisation $\Phi$ derived from the fit of the template to the OFF run were assumed for this dataset. This allows us to adjust the background rate to the different observation conditions without the need to have $\gamma$-ray emission-free regions in the ON run.

\subsection{Creation of validation datasets}
In  order to validate the above described background estimation method, the background rate and source parameters achieved by applying the run-matching approach should be compared to the standard background model template method. To achieve this, we construct validation datasets consisting of observations of gamma-ray sources that are either point-like or exhibit only marginal extension, facilitating the application of both methods. For this purpose, different datasets for all regions of interest (ROI) around the sources were constructed, to estimate the accuracy of the background description using the run-matching approach. For ease of reference, these different cases have been given numbers and an overview can be found in Table \ref{tab:cases}. First, a standard analysis of all ON runs for a ROI is performed as a reference. For this purpose, all observations of the target region are identified, the background model template is fitted to each ON run and the resulting \texttt{MapDatasets} are 'stacked', which means the measured data is summed over all observations, averaged IRFs are created and only one \texttt{MapDataset} is returned for every ROI. Following, this dataset will be referred to as Case~0. Another dataset was constructed using the same approach, just with the run-matching approach, but without corrections, to estimate the background rate. This will be referred to as Case~1. One dataset with only the correction for the differences in duration was computed (Case~2) and one where both corrections were applied (Case~3). Two more datasets were constructed to estimate the influence of the systematic errors introduced due to the run-matching approach (see Section \ref{sys_errors}), they are labelled with Case~4+ and Case~4- respectively for increased and decreased background count rate.

\begin{table}[ht]
\caption{Overview over the identifiers used for the datasets in this validation.} 
\label{tab:cases} 
\centering                          
\begin{tabular}{c  | c  }       
\hline\hline  
\noalign{\smallskip}
Case & Description \\    
\noalign{\smallskip}
\hline 
\noalign{\smallskip}
    Case 0 & standard background model template analysis\\
    Case 1 & run-matching approach \\
    Case 2 & Case 1 with a duration correction \\
    Case 3 & Case 2 with a zenith angle correction \\
    Case 4+ & Case 3 with added systematic uncertainties \\
    Case 4- & Case 3 with subtracted systematic uncertainties \\
\noalign{\smallskip}
\hline   
\hline 
\end{tabular}
\end{table}

\subsection{Derivation of the correlation coefficients and validity intervals}
\begin{figure}
\includegraphics[width=9cm]{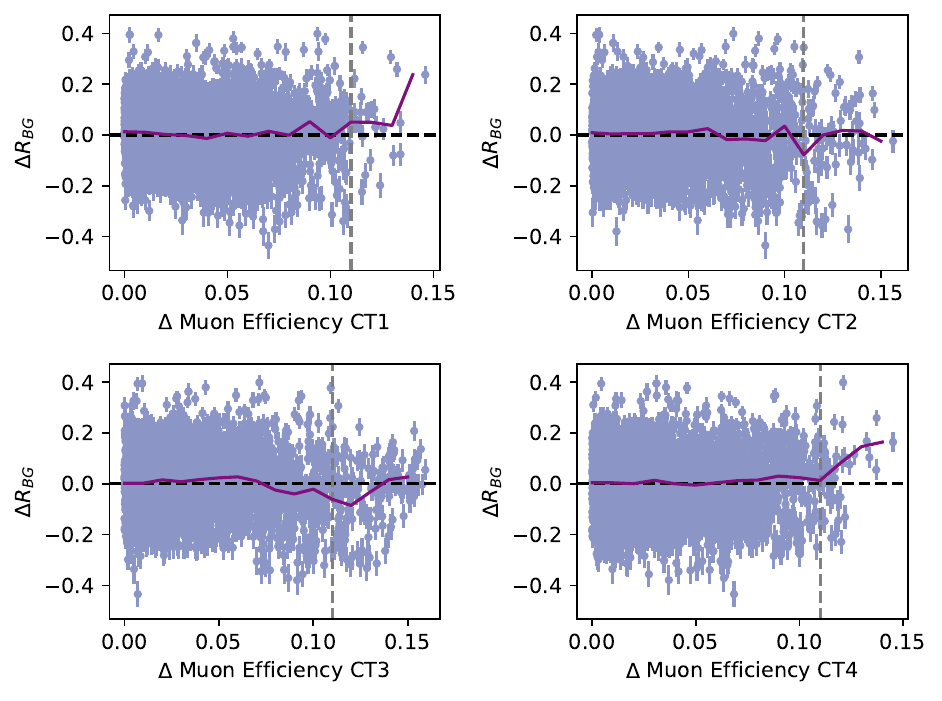}
\caption{The background rate deviation $\Delta R_{BG}$ per difference in muon efficiency for each run pair in set~3 (see Table \ref{tab:datasets} for more information on the dataset). The case of no deviation between the background rates is depicted by the grey dashed line, the mean of $\Delta R_{BG}$ is depicted by the purple line.}
\label{fig:valid_range}
\end{figure}

The influence of the respective matching parameters on the background rate is identified by analising archival H.E.S.S. data taken on the $\gamma$-ray source PKS~2155$-$304 using the direct application of the background model template (Case~0). 
\\
PKS~2155$-$304 was chosen as a test region, as it has been continuously monitored since the commissioning of the first H.E.S.S. telescope and a large amount of data, with varying observation conditions has been acquired. Additionally, the $\gamma$-rays in this ROI are contained in a small, well-known region, resulting in a small uncertainty in the background rate.
\\
The background model template was fitted to each observation in this dataset (see Section \ref{background-fit}), and the number of background events was estimated. The data was then split into three groups depending on the hardware configuration of the telescope array. A total of 791 observations over the three hardware phases was used. A Pearson correlation coefficient between the different matching parameters and the background rate was then computed. The results can be seen in Table \ref{tab:params_deviations}.
\\
To estimate the valid parameter range for every matching parameter, a comparison of the background rate estimated for Case~0 and the run matching approach (Case~3) is carried out for a large number of observations. For this purpose all observation pairs from the sets of observations on PKS~2155$-$304, indicated in Table \ref{tab:systematic}, are used.
\\
For each of these observations, all possible OFF runs are identified. Then, the background rate of all pairs is computed for Case~0 and Case~3, and the deviation of these is calculated as:
\begin{equation}
\label{background_rate}
   \Delta R_{BG} = \frac{R_{BG, 0} - R_{BG, 3}}{R_{BG, 3}}
\end{equation}
with $R_{BG, 0}$ the background rate of the dataset for Case~0 and $R_{BG, 3}$ the background rate for Case~3. Additionally, the deviation between all matching parameters of ON and OFF run as $\Delta x = (x_\text{on} - x_\text{off})/x_\text{on}$, with $x$ a matching parameter, for each observation pair, is calculated. We then compute the mean background rate deviation $\overline{\Delta R_\text{BG}}$ per $\Delta x$, and define the valid parameter range as the $\Delta x$ at which $|\overline{\Delta R_\text{BG}}| > 10\%$. This computation is performed individually for all four telescopes, and the smallest value is identified as the upper bound of the validity range. A visualisation of the distribution of $\overline{\Delta R_{BG}}$ and its mean for the Muon efficiency can be seen in Figure \ref{fig:valid_range}. 

\section{Systematic errors}\label{sys_errors}
In order to estimate the systematic uncertainties introduced by employing the run-matching approach, a direct comparison between the background rate for Case~0 and Case~3 is made. For this purpose, a set of observations on a well-known target region is selected. For each of these observations, all possible OFF runs are identified. Then, the background rate of all pairs is computed for Case~0 and Case~3, and the deviation of these is calculated as indicated in Equation \ref{background_rate}. This deviation, a measure of the systematic shift, is computed for all run pairs and the results are grouped according to the fractional run difference of the respective OFF run. Despite this comparison being a good estimation for the systematic uncertainty introduced by the run matching approach, it is limited by the available statistics, and a coarse bin size of $\Delta f = 0.1$ is chosen. Subsequently, the standard deviation of the background deviation was computed in each bin with more than 10 entries, to ensure sufficient statistics for a stable result. The standard deviation was then used as systematic uncertainty on the background count rate for a run pair with the respective fraction run deviaton.
\\
Because of strong variations in the optical efficiency of the telescopes, the systematic uncertainty can vary for observations obtained at different times. To account for this effect, the systematic uncertainties should be computed for each analysis respectively, by selecting observations recorded in a short time-span around the recording of the ON runs used for the source analysis. In this study, ON runs from three different time periods are used, therefore three different sets of observations for the estimation of the systematic uncertainties are computed. For all three sets observations on the source PKS~2155$-$304, excluding the observations which are part of the public data release, were used. Further properties of these sets, as well as which source analyses they were used for, can be seen in Table \ref{tab:systematic}.

\begin{table*}[ht]
\caption{Properties of the sets of observations from which the systematic uncertainties were estimated and the corresponding source analysis they were used for.} 
\label{tab:systematic} 
\centering                          
\begin{tabular}{c  | c c c c }       
\hline\hline  
\noalign{\smallskip}
 Dataset & Optical phase & ON runs & OFF runs & Source analysis \\    
\noalign{\smallskip}
\hline 
\noalign{\smallskip}
   Set~1  & 1b & 49 & 284 & Crab Nebula, MSH~15$-$5\textit{2}, RX~J1713.7$-$3946 \\
   Set~2  & 1c & 38 & 2170  & PKS~2155$-$304, Sculptor Dwarf Galaxy\\
   Set~3  & 2c2 & 27 & 631 & Reticulum \RomanNumeralCaps 2, Tucana \RomanNumeralCaps 2\\
\noalign{\smallskip}
\hline   
\hline 
\end{tabular}
\end{table*}

\begin{figure}
\includegraphics[width=9cm]{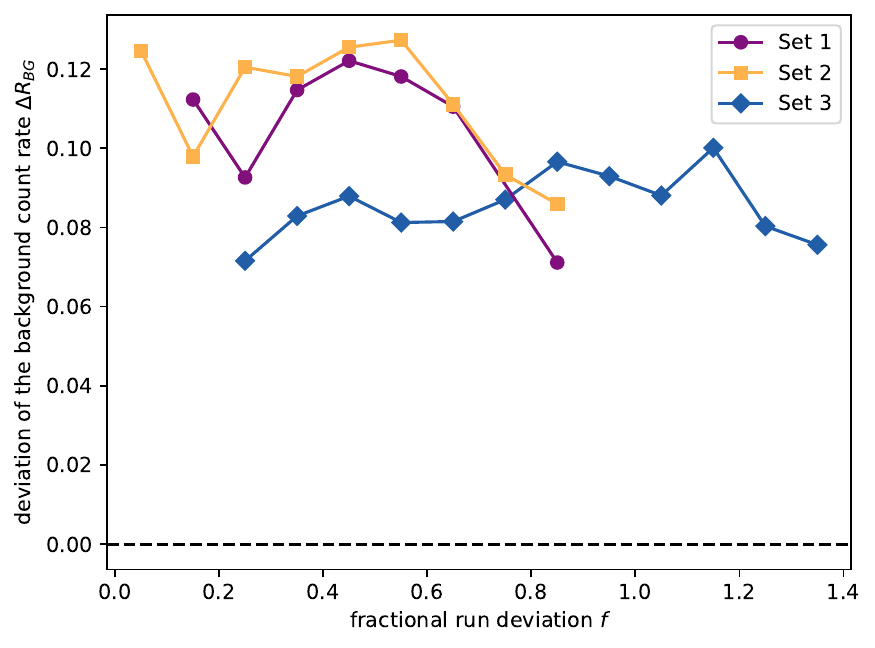}
\caption{The systematic uncertainties on the background count rate introduced by using a run-matching approach for different fractional run deviations $f$ for all datasets. The error bars represent the standard deviation of all pairs in the respective $f$ bin. A description of the respective sets can be found in Table \ref{tab:systematic}}
\label{fig:systematics}
\end{figure}

A visualisation of the systematic shift for all three sets can be seen in Figure \ref{fig:systematics}. The systematic errors for set~1 and set~2 are comparable, whilst the errors for set~3 are marginally smaller. This is most likely caused by a camera update in 2016 \citep{hess1u}. Set~3 also extends to higher fractional run deviations $f$, since a larger amount of observations on extragalactic sources was acquired in this time period. 
\\
The influence of this systematic shift of the background rate on the source parameters is then estimated by computing two additional datasets for each test region. For the computation of these datasets, the fractional run difference of every run pair was used to identify the systematic shift expected for this pair and the background counts per pixel were then increased and decreased using the corresponding systematic factor. These datasets are identified as Case~4+, for the increased background rate and Case~4- for the decreased background rate.

\section{Validation}\label{validation}
To validate the run-matching approach, data from the H.E.S.S. public dataset release \citep{data-release} was analysed (Case~3) and the results compared to an analysis using the background model template (Case~0), as well as the results reported in \citet{bkg_template}. Additionally, this work uses observations of sky regions devoid of $\gamma$-ray emission (which will be referred to as empty-field observations hereafter) acquired with the H.E.S.S. telescope array. We note that the archival data used for the construction of the background model template, as well as the OFF runs used for these analyses, and the observations of the empty-field observations are proprietary to the H.E.S.S. Collaboration and not publicly available.
\\
The public data release only contains data passing a tight quality selection (for more information see \cite{sys_index}). Therefore all observations that are part of the data release can be used for this validation study. The empty-field observations were filtered to only contain runs taken under good atmospheric conditions. The properties of these datasets are listed in Table \ref{tab:datasets}.
\\
The spectral and morphological properties of the $\gamma$-ray emission from all sources contained in the public data release (see Table \ref{tab:datasets}) were acquired by performing a 3-dimensional fit of a spectromorphological model to the dataset. A `stacked' analysis was performed using \texttt{Gammapy}, i.e. the data is summed over all observations, and the likelihood minimisation of the model fit parameters is carried out over averaged IRFs.
\\
All datasets were prepared using a spatial pixel size of $0.02^\circ$. The spatial extension of the \texttt{MapDataset}s for each analysis region was chosen such that all events recorded in the observations were included in the analysis, and the energy axis for each dataset was chosen to be logarithmically spaced, with eight bins per decade. 
\\
\begin{table*}[ht]
\caption{The analysis datasets used in this study.} 
\label{tab:datasets} 
\centering  
\begin{threeparttable}
\begin{tabular}{c c | c c c c c }       
\hline\hline  
\noalign{\smallskip}
Target & Type & Extension & Runs & Observation time [h] & Zenith angle [$^\circ$] & $\overline{f}$  \\    
\noalign{\smallskip}
\hline 
\noalign{\smallskip}
    Reticulum \RomanNumeralCaps 2\tnote{d} & dSph & None       & 16 & 7.0  & $34^\circ - 50^\circ$ & 0.24 \\
    Tucana \RomanNumeralCaps 2\tnote{d}    & dSph & None       & 14 & 6.2  & $35^\circ - 38^\circ$ & 0.28 \\
    Sculptor Dwarf Galaxy\tnote{e}                      & Sph & None       & 22 & 10.0 & $11^\circ - 19^\circ$ & 0.22  \\
    Crab Nebula\tnote{b}                  & PWN  & Point-like & 4  & 1.9  & $45^\circ - 49^\circ$ & 0.35   \\
    PKS~2155$-$304 (steady)\tnote{a,b}       & AGN  & Point-like & 6  & 2.8  & $23^\circ - 37^\circ$ & 0.89  \\
    MSH~15$-$5\textit{2}\tnote{b}          & PWN  & Extended   & 15 & 6.8  & $36^\circ - 39^\circ$ & 0.22  \\
    RX~J1713.7$-$3946\tnote{b}             & SNR  & Extended   & 15 & 7.0  & $16^\circ - 26^\circ$ & 0.58  \\
    RX~J1713.7$-$3946 (large)\tnote{c}    & SNR  & Extended   & 153& 51.0 & $17^\circ - 59^\circ$ & 0.57  \\
\noalign{\smallskip}
\hline   
\hline 
\end{tabular}
\tablefoot{The source extension noted in this table does not necessarily match the best description found in the literature, but rather the description used in \citet{bkg_template}, as well as this analysis.}
\begin{tablenotes}
            \item[a] Observations of PKS~2155$-$304 taken outside of periods when the source undergoes $\gamma$-ray outbursts
            \item[b] This dataset is part of the first public data release of the H.E.S.S. collaboration \citep{data-release}
            \item[c] More information on this dataset can be found in \citet{rxj}
            \item[d] More information on this dataset can be found in \citet{dwarf1}
            \item[e] More information on this dataset can be found in \citet{dwarf2}
        \end{tablenotes}
\end{threeparttable}
\end{table*}

\subsection{Validation with empty-field observations}
The quality of the background estimation was assessed via the \citet{lima} significance distribution. For an empty sky region, a Gaussian distribution centred around $\mu = 0$, with a standard deviation of $\sigma = 1$, is expected, since the fluctuations of the background counts are Poisson distributed.  
\\
To test for the correct description of the background, three datasets comprised of observations centred on the empty-filed regions, namely regions around the dwarf spheroidal galaxies Reticulum \RomanNumeralCaps 2, Tucana \RomanNumeralCaps 2 and Sculptor Dwarf Galaxy, were examined. These regions were chosen because no significant $\gamma$-ray emission from the sources, as well as within a $4^\circ$ region around the sources, has been observed with H.E.S.S. \citep{dwarf1, dwarf2}.Therefore, the regions can be used for an estimation of the background rate without contamination from a mismodelled $\gamma$-ray source.

\begin{figure}
\includegraphics[width=9cm]{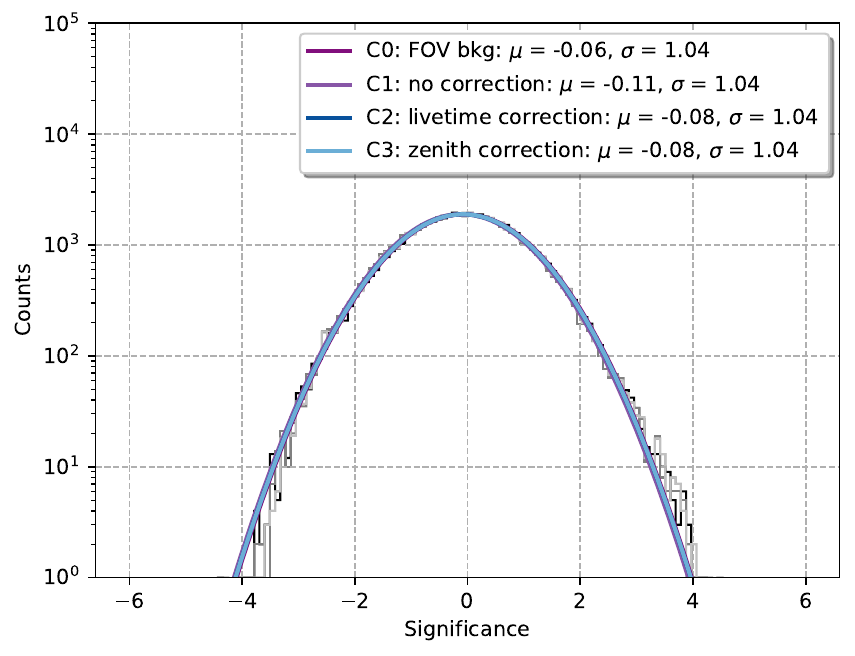}
\caption{Li\&Ma significance distribution of the observations in the region around the dwarf spheroidal galaxy Reticulum \RomanNumeralCaps 2. Depicted are the distributions for four different estimates of the background, see Table \ref{tab:background_hist}. Each of the distributions is shown as a histogram in grey, while a Gaussian fit to each distribution is shown by the coloured lines.}
\label{fig:sig-dist-tucana}
\end{figure}

\begin{table}[ht]
\caption{Gaussian distribution of the Li\&Ma significance values of the background events derived from the empty-field observations.} 
\label{tab:background_hist} 
\centering                          
\begin{tabular}{c  || c c | c c| c c |c c }       
\hline\hline  
\noalign{\smallskip}
Case & \multicolumn{2}{|c|}{Reticulum \RomanNumeralCaps 2}  & \multicolumn{2}{|c|}{Tucana \RomanNumeralCaps 2} & \multicolumn{2}{|c|}{Sculptor} \\    
\noalign{\smallskip}
\hline 
\noalign{\smallskip}
            & $\mu$ & $\sigma$ & $\mu$ & $\sigma$ & $\mu$ & $\sigma$ \\
    Case 0  & $-0.06$& $1.04$ & $-0.05$& $1.01$ & $-0.10$& $1.04$ \\
    Case 1  & $-0.11$& $1.04$ & $-0.08$& $1.05$ & $-0.15$& $1.03$\\
    Case 2  & $-0.08$ & $1.04$ & $-0.07$ & $1.06$ & $-0.14$ & $1.03$ \\
    Case 3  & $-0.08$ & $1.04$ & $-0.06$ & $1.06$ & $-0.14$ & $1.03$ \\
    Case 4+ & $-0.30$ & $1.03$ & $-0.24$ & $1.04$ & $-0.42$ & $1.01$ \\
    Case 4- & $ 0.15$ & $1.06$ & $ 0.12$ & $1.07$ & $ 0.15$ & $1.07$ \\
\noalign{\smallskip}
\hline   
\hline 
\end{tabular}
\tablefoot{The mean $\mu$ and standard deviation $\sigma$ are derived from a Gaussian fit to the distribution of the Li\&Ma significance values of the background events. The uncertainties are of the order of $10^{-4}$ to $10^{-5}$ on all fit parameters.}
\end{table}
For all regions, the significance of the number of events passing selection cuts in excess of the background prediction was computed. The correlation radius used for the construction of the significance maps is $0.06^\circ$, corresponding approximately to the point-spread function of H.E.S.S., for all empty-field observations. A Gaussian model is fitted to each significance distribution with the fit results for the different regions given in Table \ref{tab:background_hist}. An example distribution for the region around the dwarf spheroidal galaxy Tucana \RomanNumeralCaps 2, and the corresponding Gaussian fit can also be seen in Figure \ref{fig:sig-dist-tucana}.

\begin{figure}
\includegraphics[width=9cm]{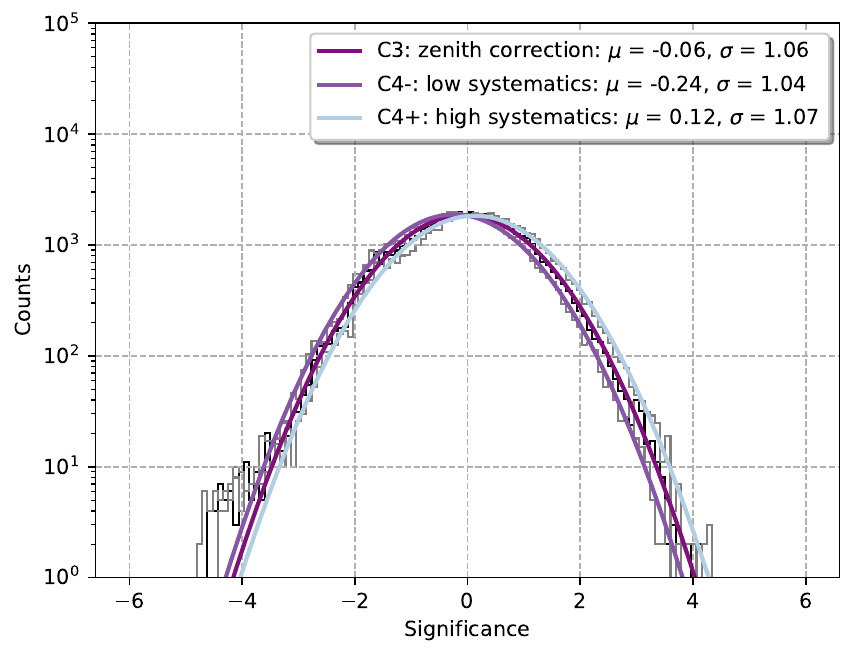}
\caption{The Li\&Ma significance distributions of background counts in the region around the dwarf spheroidal galaxy Tucana \RomanNumeralCaps 2 for the different background estimation methods.}
\label{fig:systematics_example}
\end{figure}

For Case~0, all three regions show a distribution centred around zero and a standard deviation of approximately 1.0, confirming that no $\gamma$-ray sources are present. The mean of the distributions for Case~1, indicates a over-prediction of cosmic-ray background of up to 20\%. This over-prediction is slightly decreased if a duration correction is applied (Case~2), and further reduced once the zenith correction is applied for all three empty-field regions.  
\\
The Gaussian fit to the significance histograms of these datasets shows that the nominal value derived for the Case~0 datasets is included in the range covered by the systematic error for all datasets. An example for the shift on the significance distribution caused by the inclusion of the systematic errors can be seen in Figure \ref{fig:systematics_example}.
\\
We also tested the validity of the background estimation method in different energy ranges, using again the three empty-field data sets. For this purpose, the data was divided into 8 logarithmically spaced energy bins, from $0.1\,$TeV to $10\,$TeV. This binning was chosen so that each bin included two energy bins of the initial dataset. The data above 10 TeV were not included in this comparison due to insufficient gamma-ray statistics. We find that for the three data sets, a Gaussian fit to the significance distribution yields comparable results between the Case~0 and Case~3 datasets in all energy bins. The mean and standard deviation for the Gaussian fits in energy bands are quoted in Tables \ref{tab:enedep_reticulum} - \ref{tab:enedep_tucana} and a visual example for the region around Sculptor can be seen in Figure \ref{fig:sigdist_enedep}. 
\\
To check for variation in background counts as a function of energy due to the run matching, we computed the background counts per energy bin for all empty-field region observations were computed for Case~0 ($N_{BKG}^0$) and Case~3 ($N_{BKG}^3$). Then, the ratio between these counts $N_{BKG}^3/N_{BKG}^0$ was computed per energy bin. We find that the number of background counts for the Case~3 datasets is slightly underestimated for lower energies and slightly overestimated for higher energies compared to the number of counts derived for the Case~0 datasets. This deviation is however found to be below $6\%$ for all energies and can be seen in Figure \ref{fig:bkg_count_rate_enedep}.

\subsection{Public data release data}
After verifying the background prediction in an empty sky region, the best-fit values for a source analysis should be verified between the different background estimation techniques. 
\\
For this purpose, data from the public data release of H.E.S.S. \citep{data-release} was analysed for both background estimation methods, and the results are compared to those derived in \citet{bkg_template}. As a spectral model for all datasets, a simple power-law was chosen and in all cases, the flux normalisation $\text{N}_0$ at a reference energy $E_0$ and the spectral index $\Gamma$ were used as fit parameters. The power-law model is defined as:
\begin{equation}
\label{powerlaw}
    \frac{\text{dN}}{\text{dE}} = \text{N}_0 \cdot \left(\frac{\text{E}}{\text{E}_0}\right)^{-\Gamma}\,\,.
\end{equation}
The reference energy for all datasets was chosen to be equal to the values used in \citet{bkg_template}, for the sake of comparison, and can be found in Table \ref{tab:fit-params}.
\\
The correlation radius used for the computation of the significance maps and histograms for all datasets is $0.06^\circ$, except the dataset centred on the Crab Nebula which is $0.1^\circ$, due to the small size and limited statistics of the dataset.
\\
For the sake of comparing the best-fit results between this analysis and the results obtained in \citet{bkg_template}, the same spatial models are chosen and should not be interpreted as yielding the most accurate description of the region. The Crab Nebula, as well as PKS~2155$-$304, are described using a Point Source Model. For a more in-depth discussion of these sources, see \citet{crab_new} and \citet{pks} respectively. The pulsar wind nebula MSH~15$-$5\textit{2} (analysed in detail in \citet{msh}) is described by an elongated disk model. 
For the supernova remnant RX~J1713.7$-$3946, no pre-defined spatial model could be used because of the complicated morphology of the source. For this reason an `excess template' was constructed (for more information about the construction of this excess template see \citet{bkg_template}). More information describing the emission from RX~J1713.7$-$3946 can be found in \citet{rxj}.
\\
An additional challenge for the analysis of these datasets is that, depending on the source location, misclassified cosmic rays are not the only source of background events. Observations centred in the galactic plane will also include events from the galactic diffuse emission \citep{diffuse_emission}. For the background estimation employing the background model template, the galactic diffuse emission can partly be absorbed by increasing the normalisation of the background model template. Since, in the case of the run-matching approach, the background model template is normalised on observations outside of the galactic plane, absorption of the diffuse emission into the background is not possible. 
\\
This effect has been observed in the analysis of the datasets centred on the regions around the sources MSH~15$-$5\textit{2}, located at a galactic latitude of $b=-1.19^\circ$ and RX~J1713.7$-$3946, with a galactic latitude of $b=-0.47^\circ$. For both datasets, an excess signal across the whole FoV was detected at low energies. Whilst it is likely that the observed excess emission is galactic diffuse emission, the data used in this validation is not extensive enough to model this signal or derive any of its physical properties. Therefore, we adopted a strict energy threshold for the analysis of the regions around MSH~15$-$5\textit{2} and RX~J1713.7$-$3946, effectively excluding the energy range in which a significant influence of diffuse emission can be observed for H.E.S.S..
This energy threshold was evaluated respectively for each analysis for Case~3. For this purpose, the number of background and signal events in a source-free region in the stacked dataset was estimated. The first energy bin after which the absolute difference between the number of signal and background events is less than $10\%$ was adopted as energy threshold for the analysis. For the analysis of both MSH~15$-$5\textit{2} and RX~J1713.7$-$3946, this resulted in an energy threshold of $560\,$GeV.
\begin{figure*}
\begin{minipage}[b]{.49\textwidth}
\includegraphics[width=\textwidth]{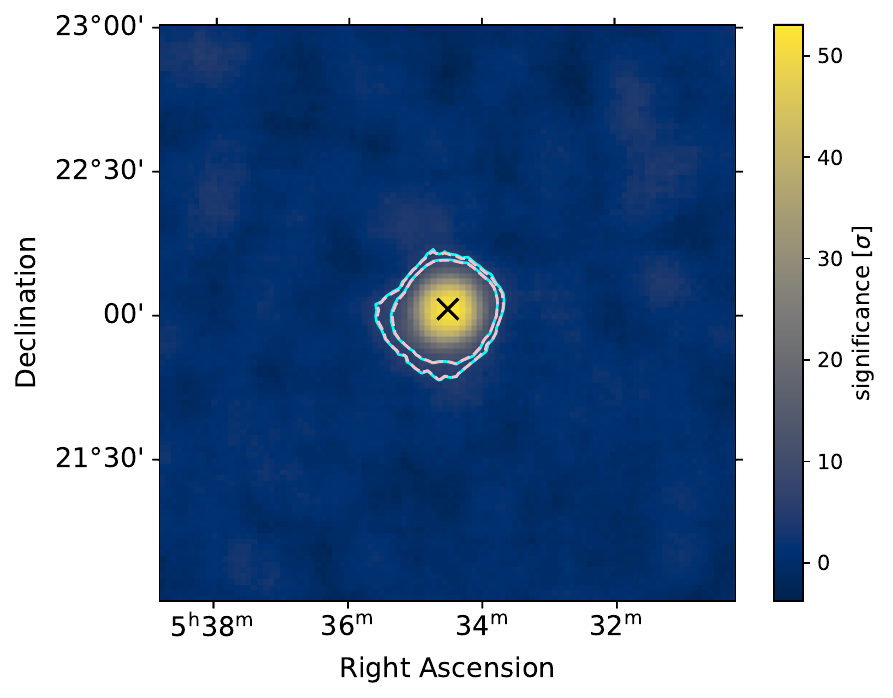}
\end{minipage}\qquad
\begin{minipage}[b]{.49\textwidth}
\includegraphics[width=\textwidth]{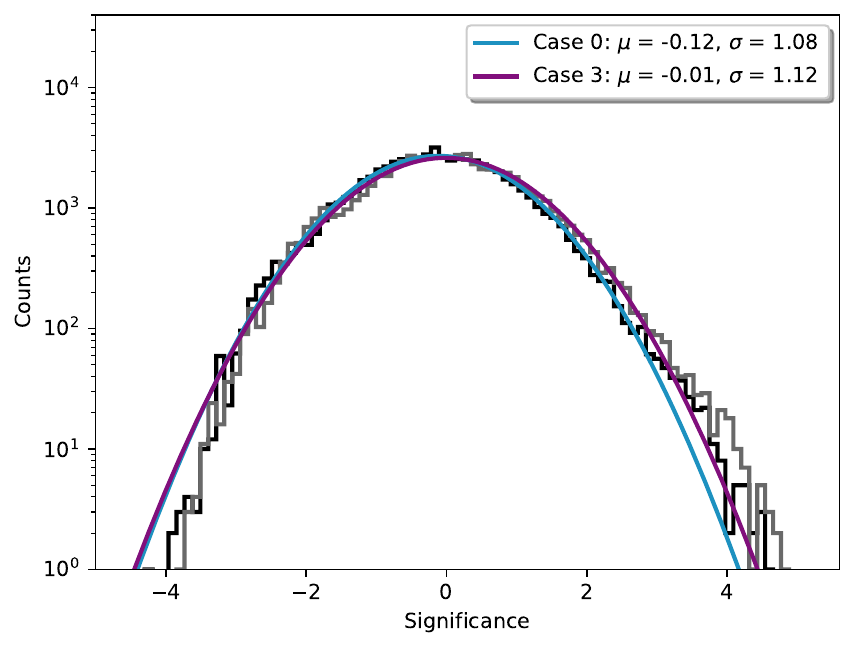}
\end{minipage}
\caption{Left: Li\&Ma significance map of the region around the Crab Nebula as seen with the H.E.S.S. telescopes, with $5\,\sigma$ and $8\,\sigma$ contours. The contours from case 0 are depicted in blue, while the contours from case 3 are depicted in pink. The best-fit position is indicated by the black cross. Right: significance distribution of the background in the FoV around the Crab Nebula. }
\label{fig:crab_sigmap}%
\end{figure*}
Additionally to the increased energy thresholds, a second modification of the background estimation needs to be included for the datasets for MSH~15$-$5\textit{2} and RX~J1713.7$-$3946. Because they were taken in 2004, shortly after the commissioning of the H.E.S.S. array, the optical phase is very short, due to fast system degradation, and few observations have been taken outside of the galactic plane in this time period. For this reason, no sufficient amount of OFF runs can be found within this optical phase and the first two optical phases have been combined.
\\
The best-fit parameters for the datasets created using the run-matching approach for all sources can be found in Table \ref{tab:fit-params}. Additionally, the best-fit values derived in the previous analysis, as well as the results derived from the Case~0 datasets are also indicated in the table for comparison.
\\
A visual comparison of the fit parameters for all sources can be found Appendix \ref{appendix:fit_results}.

\begin{table*}[ht]
\caption{Gaussian distribution of the Li\&Ma significance values of the background events derived for each of the datasets in the H.E.S.S. public data release.} 
\label{tab:background_hist_dataset} 
\centering                          
\begin{tabular}{c  || c c |c c| c c |c c | c c}       
\hline\hline  
\noalign{\smallskip}
Case & \multicolumn{2}{|c|}{Crab Nebula} &  \multicolumn{2}{|c|}{PKS~2155$-$304} & \multicolumn{2}{|c|}{MSH~15$-$5\textit{2}} & \multicolumn{2}{|c|}{RX~J1713.7$-$3946} & \multicolumn{2}{|c|}{RX~J1713.7$-$3946 (large)} \\    
\noalign{\smallskip}
\hline 
\noalign{\smallskip}
& $\mu$ & $\sigma$ & $\mu$ & $\sigma$ & $\mu$ & $\sigma$ & $\mu$ & $\sigma$ & $\mu$ & $\sigma$\\
    Case 0  & $-0.12 $& $1.08$ & $-0.11$& $1.04$ & $-0.16 $& $1.04 $ & $-0.26 $& $1.02 $ & $-0.20 $& $1.05$\\
    Case 3  & $-0.01$& $1.12$ & $-0.17$ & $1.03$ & $-0.29$ & $1.02$ & $-0.29$ & $1.03$ & $-0.18$ & $1.05$\\
    Case 4+ & $-0.34$ & $1.18$ & $-0.34$ & $1.06$ & $-0.68$ & $1.05$ & $-0.56$ & $1.07$ & $-0.81$ & $1.11$ \\
    Case 4- & $ 0.36$ & $1.08$ & $ 0.01$ & $1.02$ & $ 0.14$ & $1.00$ & $ 0.00$ & $1.00$ & $-0.51$ & $1.23$ \\  
\noalign{\smallskip}
\hline   
\hline 
\end{tabular}
\tablefoot{The mean $\mu$ and standard deviation $\sigma$ derived by fitting a Gaussian to the distribution of the Li\&Ma significance values of the background events derived for each of the datasets in the H.E.S.S. public data release. The uncertainties on $\mu$ and $\sigma$ are of the order of $10^{-4}$ to $10^{-5}$ for all regions.}
\end{table*}

\subsubsection{Point-like sources}
The comparison of the best-fit parameters for the analysis of the $\gamma$-ray emission from the Crab Nebula and PKS~2155$-$304 can be found in Figure \ref{fig:params-crab} and Figure \ref{fig:params-pks} respectively. 
\\
The left panel of Figure \ref{fig:crab_sigmap} shows a significance map for the region around the Crab Nebula. Indicated in the significance map are the $5\,\sigma$ and $8\,\sigma$ contours for Case~3 in dashed pink lines, and the contours for Case~0 as solid blue lines. The contours show good agreement. 
\\
The right panel of Figure \ref{fig:crab_sigmap} shows the distribution of significance entries in the respective maps. A region of $0.5\,^\circ$ radius around the source has been excluded to avoid contamination of residual $\gamma$-ray emission from the source. The significance map and histogram of the region around PKS~2155$-$304 can be found in Figure \ref{fig:pks_sigmap}. The distribution of the background counts shows a shift between the datasets for Case~0 and Case~3, indicating a slight over-prediction of the background rate for the Case~3 dataset. This shift is however within the expected range derived from a study of the influence of the systematics (see Table \ref{tab:background_hist_dataset})
\\
The best-fit results of the likelihood minimisation for the Crab Nebula can be seen in Figure \ref{fig:params-crab}. All parameters except the best-fit position are comparable within statistical errors with the results derived in \citet{bkg_template}. The comparability of the Right Ascension between Case~0 and Case~3 suggests that a change in pointing reconstruction might be responsible for this deviation. A similar deviation can be observed for the best-fit parameters of PKS~2155$-$304 (see Figure \ref{fig:params-pks}), as well as a slight decrease in flux normalisation compared to the results derived in \citet{bkg_template}. For both sets of SEDs, depicted in Figure \ref{fig:sed_crab} and Figure \ref{fig:sed_pks}, good agreement is seen. Only in one energy bin, around $3\,$TeV is the deviation above $1\,\sigma$ between the SED derived from the Crab Nebula. A reason for this deviation could be a change in the binning of the background model template that has been incorporated since the previous publication from \citet{bkg_template}. 
\\

\begin{figure*}
\centering
\includegraphics[width=15cm]{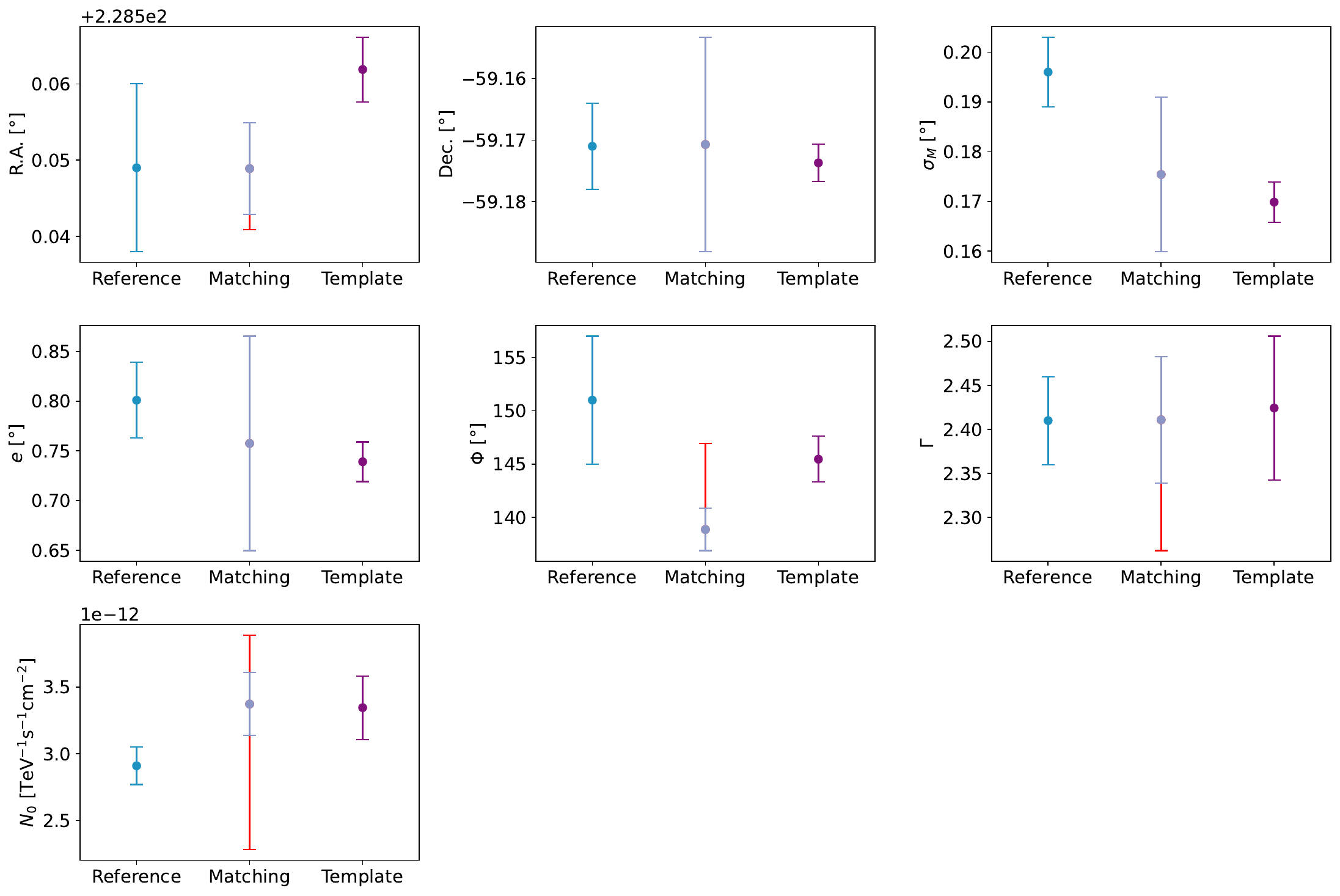}
\caption{Comparison between the best-fit values for all parameters of the elongated model used to describe the emission from MSH~15$-$5\textit{2}. The reference values where taken from \citet{bkg_template}, Matching refers to the values derived by using the run-matching approach for the background estimation, while template refers to the background model template. The systematics introduced due to the run-matching approach are indicated in red.}
\label{fig:params-msh}
\end{figure*}

\subsubsection{MSH~15$-$5\textit{2}}
A significance map of the region around MSH~15$-$5\textit{2}, as well as the distribution of background can be seen in Figure \ref{fig:msh_sigmap}. In this region a shift of the background distribution towards a smaller number of cosmic-ray signals can be observed. This shift is most likely caused by a lack of OFF runs in this optical phase, making it necessary to choose OFF runs from the next optical phase. It is, therefore, possible that the time-span between ON and OFF run is larger than $4\,$years resulting in a high probability that the optical efficiency of the telescopes between both observations differs substantially. The shift can, however, be accounted for by the systematic uncertainty introduced by the run matching approach.  
\\
The best-fit values for all parameters agree within the error (see Figure \ref{fig:params-msh}). The strong dependence of the source extension on the correct estimation of the background rate can be well observed by the large systematic errors introduced by the systematic uncertainty due to the run-matching. The SED derived for this source shows a good agreement within the errors with the SED computed in \citet{bkg_template} (Figure \ref{fig:sed_msh}).

\subsubsection{RX~J1713.7$-$3946}
\begin{figure}
\includegraphics[width=9cm]{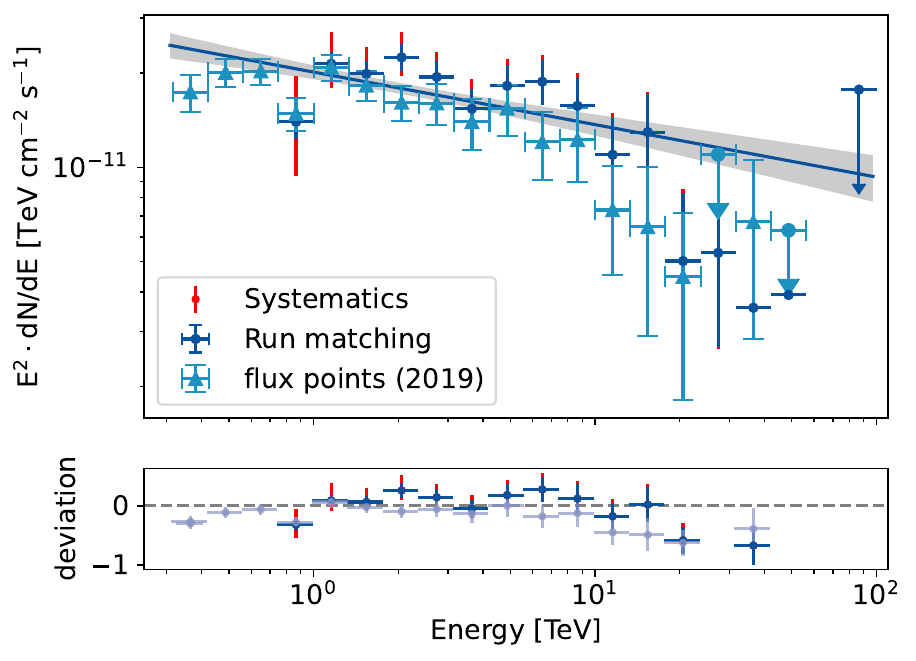}
\caption{The SED of RX~J1713.7$-$3946 derived from this analysis compared to the SED derived in \citet{bkg_template}. The systematics added by the run-matching approach are indicated in red in the upper panel. The lower panel shows the deviation between the SED derived in both analyses and the best-fit model derived from the Case~3 dataset.}
\label{fig:fluxpoints-msh}
\end{figure}

Figure \ref{fig:rxj_sigmap} shows the significance map and distribution for the region around  RX~J1713.7$-$3946. Good agreement between the background rate estimated from Case~0 and Case~3 is again seen.
\\
The best-fit parameters can be seen in Figure \ref{fig:params-rxj}. The spectral index and flux normalisation for RX~J1713.7$-$3946 agree with the results of the likelihood minimisation within the error. A comparison of the SED from RX~J1713.7$-$3946 can be seen in Figure \ref{fig:fluxpoints-msh}. The upper panel of the Figure shows the SED derived from the run-matching approach, compared to the SED derived in \citet{bkg_template}, while the lower panel shows the deviation between both sets of SED. Again, a good agreement can be observed.
\\
To verify that this background estimation is also stable for a large amount of observations, a dataset containing 53.4 hours of observations centred on RX~J1713.7$-$3946, was analysed. 
The results of this analysis show a good agreement between the Case~0 and Case~3. Since this dataset is not part of the public dataset release, the SED and best-fit parameters will not be presented in this work, but a more detailed analysis of this region using the same dataset can be found in \citet{rxj}.

\section{Conclusions}
In this work, we present a method to combine the classical ON/OFF background estimation technique used by IACT arrays with a 3D background model template. This combination of techniques allows us to remove a major restriction of each method. Since the 3D background model template has been created from a large amount of OFF runs, it is robust and not subject to large statistical uncertainties. However, this method requires source-free regions in the FoV of the observation in order to normalise the background estimation for varying observation conditions. The classical ON/OFF background estimation does not require a source-free region, is however very sensitive to variations in observation conditions. 
\\
\\
Even though we are able to achieve good agreement with the previously published results for all sources, the datasets used here already give an indication of the limitations of this technique. To achieve a comparable background rate between ON and OFF run, both runs need to be chosen in a period with comparable optical efficiency. Due to a limited amount of available OFF runs in these periods, large systematic errors can be introduced on the source parameters. This effect could be mitigated by constructing a background model template suitable for long-term hardware conditions, with corrections to the normalisation designed to account for short term variations, and therefore expand the pool of viable OFF runs for these periods. 
\\
While this study presents a detailed analysis of the systematic uncertainty on the run-matching approach, an additional source of uncertainties, which has not been examined in detail, is introduced because of the stacking of the observations. This method of combining the observations for the analysis can produce a slight gradient in the number of background counts over the FoV. While this effect is small, it should nevertheless be kept in mind when interpreting the results achieved by this method and becomes particularly relevant for large, extended sources. 
\\
An additional problem of this framework is how to treat the galactic diffuse emission. In this work, we have increased the energy threshold to exclude the influence of the diffuse emission, this is however not ideal. A better approach would be to construct an additional spectromorphological template from interstellar gas tracers.
\\
While these disadvantages show that the run-matching approach presented here cannot compete with the 3D background model template method in regions containing few sources with a small extension, the non-dependence on source-free regions in the observation is a major advantage in sky regions with many sources or for the detection of diffuse extended structures, which could not be observed using the background model template alone. The presented background estimation technique is an extension of the background estimation using a 3D background model template, offering an alternative for the analysis of regions filled with significant emission and other techniques, which are subject to bigger statistical and systematic fluctuations, would need to be used. This application is vital considering that there are currently no Water Cherenkov Detectors observing the southern $\gamma$-ray sky, restricting the observable source population to structures that can be detected with the comparatively small FoV of the H.E.S.S. array. It also has the potential to increase the population of sources observable with CTA, opening up the possibility of using the superior angular resolution to expand the knowledge on large extended structures previously uniquely detected by the Water Cherenkov Detectors.

\begin{acknowledgements}
This work made use of data supplied by the H.E.S.S. Collaboration. We would like to thank the H.E.S.S. Collaboration, especially Stefan Wagner, Spokesperson of the Collaboration, as well as Nukri Komin, chair of the Collaboration Board, and Markus Boettcher, Chair of the Publication Board, for allowing us to use the data presented in this publication. 
\\
We also gratefully acknowledge the computing resources provided by the Erlangen
Regional Computing Center (RRZE).
\\

This work was supported by the German
      \emph{Deut\-sche For\-schungs\-ge\-mein\-schaft, DFG\/} project
      number 452934793.
\end{acknowledgements}

\bibliographystyle{aa} 
\bibliography{mybibliography.bib}

\begin{appendix}

\section{Zenith angle correction}\label{zenith_correction_sec}
In order to account for differences in mean zenith angle of the observation between ON and OFF run, equation \eqref{eqn:zenith_correction} is fitted to the array trigger rates in every optical phase. The resulting fit parameters are given in Table \ref{tab:zenith_correction}. 

\begin{table}[ht]
\caption{Zenith angle correction to the background rate for different optical phases.} 
\label{tab:zenith_correction} 
\centering                          
\begin{tabular}{c c | c c c c }       
\hline\hline  
\noalign{\smallskip}
Optical phase & Start date & $p_1$ & $p_2$ \\    
\noalign{\smallskip}
\hline 
\noalign{\smallskip}
   1/1b  & 01/2004 & $218.8 \pm 1.0$ & $0.92 \pm 0.02$ \\
   1c  & 07/2007 & $171.7 \pm 0.7$ & $1.10 \pm 0.06$ \\
   1c1  & 04/2010 & $174.3 \pm 1.9$ & $1.02 \pm 0.07$ \\
   1c2  & 10/2010 & $176.5 \pm 3.2$ & $0.71 \pm 0.05$ \\
   1c3  & 04/2011 & $188.2 \pm 1.9$ & $0.61 \pm 0.07$ \\
   1d  & 11/2011 & $214.9 \pm 2.7$ & $0.81 \pm 0.16$ \\
   2b0  & 01/2013 & $258.7 \pm 9.7$ & $0.73 \pm 0.21$ \\
   2b2  & 06/2013 & $250.9 \pm 12.0$ & $0.61 \pm 0.04$ \\
   2b3  & 05/2014 & $165.7 \pm 1.4$ & $0.91 \pm 0.03$ \\
   2b4  & 11/2014 & $205.2 \pm 1.6$ & $0.87 \pm 0.07$ \\
   2b5  & 08/2015 & $142.2 \pm 1.8$ & $0.84 \pm 0.19$ \\
   2c0  & 01/2017 & $380.6 \pm 17.7$ & $1.31 \pm 0.06$ \\
   2c1  & 03/2017 & $495.8 \pm 4.9$ & $1.05 \pm 0.04$ \\
   2c2  & 05/2017 & $435.0 \pm 2.3$ & $0.99 \pm 0.02$ \\
   2d3  & 10/2019 & $385.8 \pm 1.5$ & $1.08 \pm 0.01$ \\
\noalign{\smallskip}
\hline   
\hline 
\end{tabular}
\tablefoot{The fit parameters are estimated from a fit of equation \eqref{eqn:zenith_correction} to the trigger rates of the H.E.S.S. telescopes for the different optical phases.}
\end{table}


\section{Energy dependence}
\begin{figure}[h]
\includegraphics[width=9cm]{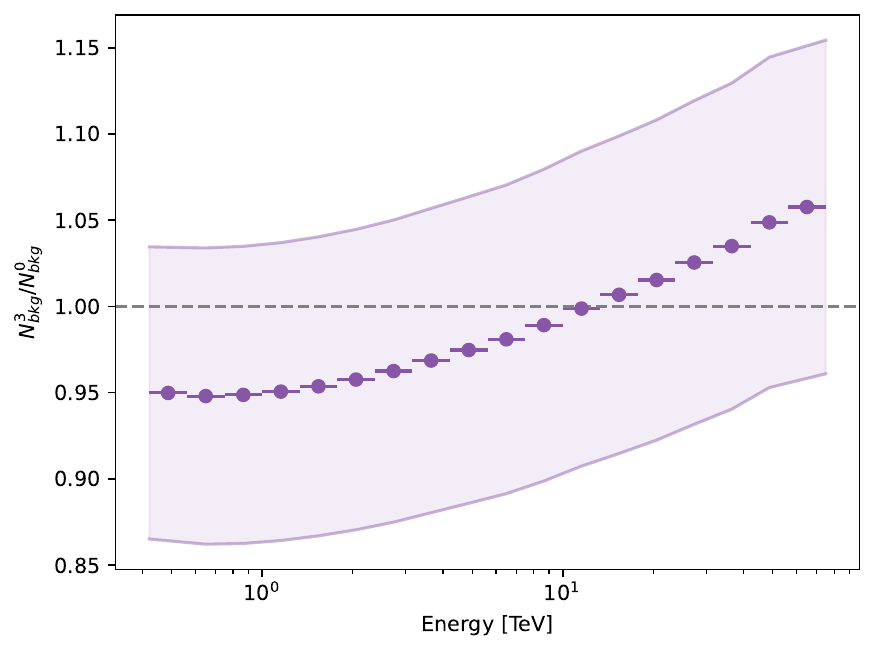}
\caption{The ratio between the number of background counts for the Case~0 ($N_{BKG}^0$) and Case~3 ($N_{BKG}^3$) datasets in energy bins. This rate was computed using all observations of the empty-field regions.}
\label{fig:bkg_count_rate_enedep}
\end{figure}

\begin{figure*}[h]
\centering
\includegraphics[width=15cm]{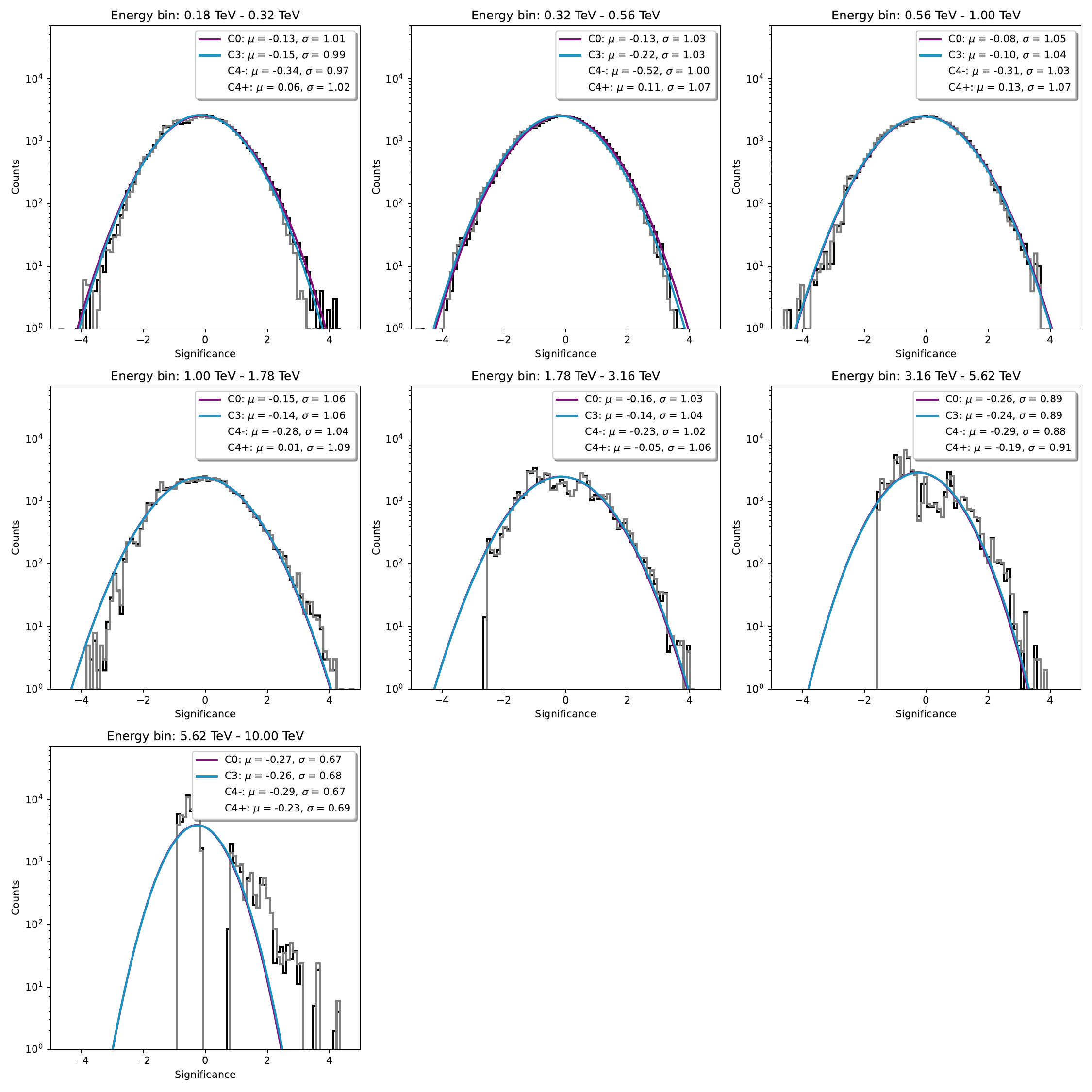}
\caption{Significance distribution for the region around Sculptor in energy bins. A Gaussian fit to the histograms for the Case~0 and Case~3 datasets is depicted by the solid lines.}
\label{fig:sigdist_enedep}
\end{figure*}

A comparison of the number of background counts for all observations used for the analysis of the empty-field regions between the Case~0 and Case~3 datasets has been made. The ratio between the background counts can be seen in Figure \ref{fig:bkg_count_rate_enedep}. The shaded band depicts the error on the number of background counts estimated from the matched pairs used for the Case~4- and Case~4+ datasets.
\\
The datasets computed for the analysis of the empty-field regions were also divided into bins, such that one bin consists of two energy bins of the original datasets. Then significance histograms were computed for all bins, and a Gaussian fit was performed. The significance histograms as well as the Gaussian fits for all bins for the dataset around Sculptor can be seen in Figure \ref{fig:sigdist_enedep} and Table \ref{tab:enedep_sculptor}. The results for the regions around Reticulum \RomanNumeralCaps 2 and Tucana \RomanNumeralCaps 2 have been summarised in Table \ref{tab:enedep_reticulum} and Table \ref{tab:enedep_tucana} respectively. 

\begin{table*}[h!]
\caption{The fit parameters of the Gaussian fit to the significance distributions estimated from the region around Reticulum \RomanNumeralCaps 2 }
\label{tab:enedep_reticulum} 
\centering                          
\begin{tabular}{c  || c c | c c| c c |c c |c c }       
\hline\hline  
\noalign{\smallskip}
Energy bin [TeV] & \multicolumn{2}{|c|}{Case 0}  & \multicolumn{2}{|c|}{Case 3} & \multicolumn{2}{|c|}{Case 4+} & \multicolumn{2}{|c|}{Case 4-} \\    
\noalign{\smallskip}
\hline 
\noalign{\smallskip}
            & $\mu$ & $\sigma$ & $\mu$ & $\sigma$ & $\mu$ & $\sigma$ & $\mu$ & $\sigma$ \\
    $0.32\,$ - $0.56\,$  & $-0.07$& $1.01$ & $-0.26$& $1.01$ & $-0.51$& $1.01 $ & $-0.00$& $1.02 $ \\
    $0.56\,$ - $1.00\,$  & $-0.03$& $1.00$ & $-0.01$& $1.02$ & $-0.25$& $1.00 $ & $ 0.25$& $1.04 $\\
    $1.00\,$ - $1.78\,$  & $-0.12$& $1.04$ & $-0.02$& $1.05$ & $-0.18$& $1.04 $ & $ 0.14$& $1.07 $ \\
    $1.78\,$ - $3.16\,$  & $-0.14$& $1.05$ & $-0.04$& $1.07$ & $-0.14$& $1.06 $ & $ 0.07$& $1.09 $ \\
    $3.16\,$ - $5.62\,$  & $-0.20$& $0.98$ & $-0.10$& $1.01$ & $-0.16$& $0.99 $ & $-0.03$& $1.02 $ \\
    $5.62\,$ - $10.00\,$ & $-0.30$& $0.78$ & $-0.24$& $0.80$ & $-0.27$& $0.80 $ & $-0.21$& $0.78 $ \\
\noalign{\smallskip}
\hline   
\hline 
\end{tabular}
\end{table*}

\begin{table*}[ht]
\caption{The fit parameters of the Gaussian fit to the significance distributions estimated from the region around Sculptor.}
\label{tab:enedep_sculptor} 
\centering                          
\begin{tabular}{c  || c c | c c| c c |c c |c c }       
\hline\hline  
\noalign{\smallskip}
Energy bin [TeV] & \multicolumn{2}{|c|}{Case 0}  & \multicolumn{2}{|c|}{Case 3} & \multicolumn{2}{|c|}{Case 4+} & \multicolumn{2}{|c|}{Case 4-} \\    
\noalign{\smallskip}
\hline 
\noalign{\smallskip}
            & $\mu$ & $\sigma$ & $\mu$ & $\sigma$ & $\mu$ & $\sigma$ & $\mu$ & $\sigma$ \\
    $0.18\,$ - $0.32\,$  & $-0.13$& $1.01$ & $-0.12$& $1.02$ & $-0.37$& $1.00 $ & $ 0.14$& $1.06 $ \\
    $0.32\,$ - $0.56\,$  & $-0.13$& $1.03$ & $-0.19$& $1.03$ & $-0.58$& $1.00 $ & $ 0.23$& $1.09 $\\
    $0.56\,$ - $1.00\,$  & $-0.08$& $1.05$ & $-0.12$& $1.04$ & $-0.39$& $1.03 $ & $ 0.18$& $1.07 $ \\
    $1.00\,$ - $1.78\,$  & $-0.15$& $1.06$ & $-0.18$& $1.06$ & $-0.36$& $1.03 $ & $-0.02$& $1.09 $ \\
    $1.78\,$ - $3.16\,$  & $-0.16$& $1.03$ & $-0.18$& $1.03$ & $-0.30$& $1.01 $ & $-0.06$& $1.06 $ \\
    $3.16\,$ - $5.62\,$ & $-0.26$& $0.89$ & $-0.28$& $0.88$ & $-0.34$& $0.87 $ & $-0.21$& $0.90 $ \\
    $5.62\,$ - $10.00\,$ & $-0.27$& $0.67$ & $-0.28$& $0.67$ & $-0.32$& $0.66 $ & $-0.25$& $0.68 $ \\
\noalign{\smallskip}
\hline   
\hline 
\end{tabular}
\end{table*}

\begin{table*}[ht]
\caption{The fit parameters of the Gaussian fit to the significance distributions estimated from the region around Tucana \RomanNumeralCaps 2 }
\label{tab:enedep_tucana} 
\centering                          
\begin{tabular}{c  || c c | c c| c c |c c |c c }       
\hline\hline  
\noalign{\smallskip}
Energy bin [TeV] & \multicolumn{2}{|c|}{Case 0}  & \multicolumn{2}{|c|}{Case 3} & \multicolumn{2}{|c|}{Case 4+} & \multicolumn{2}{|c|}{Case 4-} \\    
\noalign{\smallskip}
\hline 
\noalign{\smallskip}
            & $\mu$ & $\sigma$ & $\mu$ & $\sigma$ & $\mu$ & $\sigma$ & $\mu$ & $\sigma$ \\
    $0.32\,$ - $0.56\,$  & $-0.13$& $1.03$ & $-0.19$& $1.03$ & $-0.58$& $1.00 $ & $ 0.23$& $1.09 $ \\
    $0.56\,$ - $1.00\,$  & $-0.08$& $1.05$ & $-0.12$& $1.04$ & $-0.39$& $1.03 $ & $ 0.18$& $1.07 $\\
    $1.00\,$ - $1.78\,$  & $-0.15$& $1.06$ & $-0.18$& $1.06$ & $-0.36$& $1.03 $ & $ 0.02$& $1.09 $ \\
    $1.78\,$ - $3.16\,$  & $-0.16$& $1.03$ & $-0.18$& $1.03$ & $-0.30$& $1.01 $ & $-0.06$& $1.06 $ \\
    $3.16\,$ - $5.62\,$  & $-0.26$& $0.89$ & $-0.28$& $0.88$ & $-0.34$& $0.87 $ & $-0.21$& $0.90 $ \\
    $5.62\,$ - $10.00\,$ & $-0.27$& $0.67$ & $-0.28$& $0.67$ & $-0.32$& $0.66 $ & $-0.25$& $0.68 $ \\
\noalign{\smallskip}
\hline   
\hline 
\end{tabular}
\end{table*}

\clearpage

\section{Fit Results}
\label{appendix:fit_results}
\begin{figure*}[h]
\centering
\includegraphics[width=15cm]{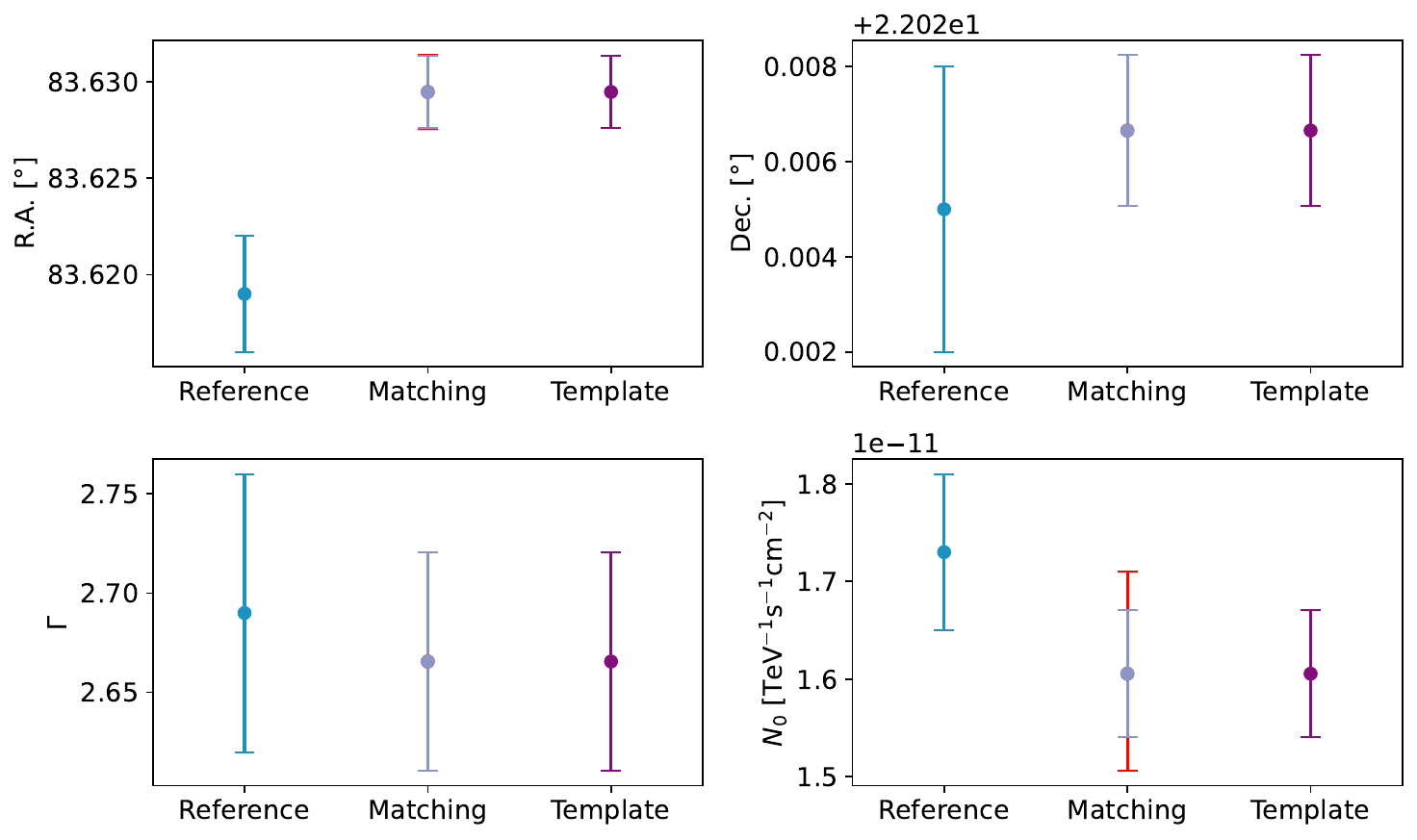}
\caption{Comparison between the best-fit values for all parameters of the point model used to describe the emission from the Crab Nebula. The reference values where taken from \citet{bkg_template}, Matching refers to the values derived by using the run-matching approach for the background estimation, while template refers to the background model template. The systematics introduced due to the run-matching approach are indicated in red.}
\label{fig:params-crab}
\end{figure*}

\begin{figure*}
\centering
\includegraphics[width=15cm]{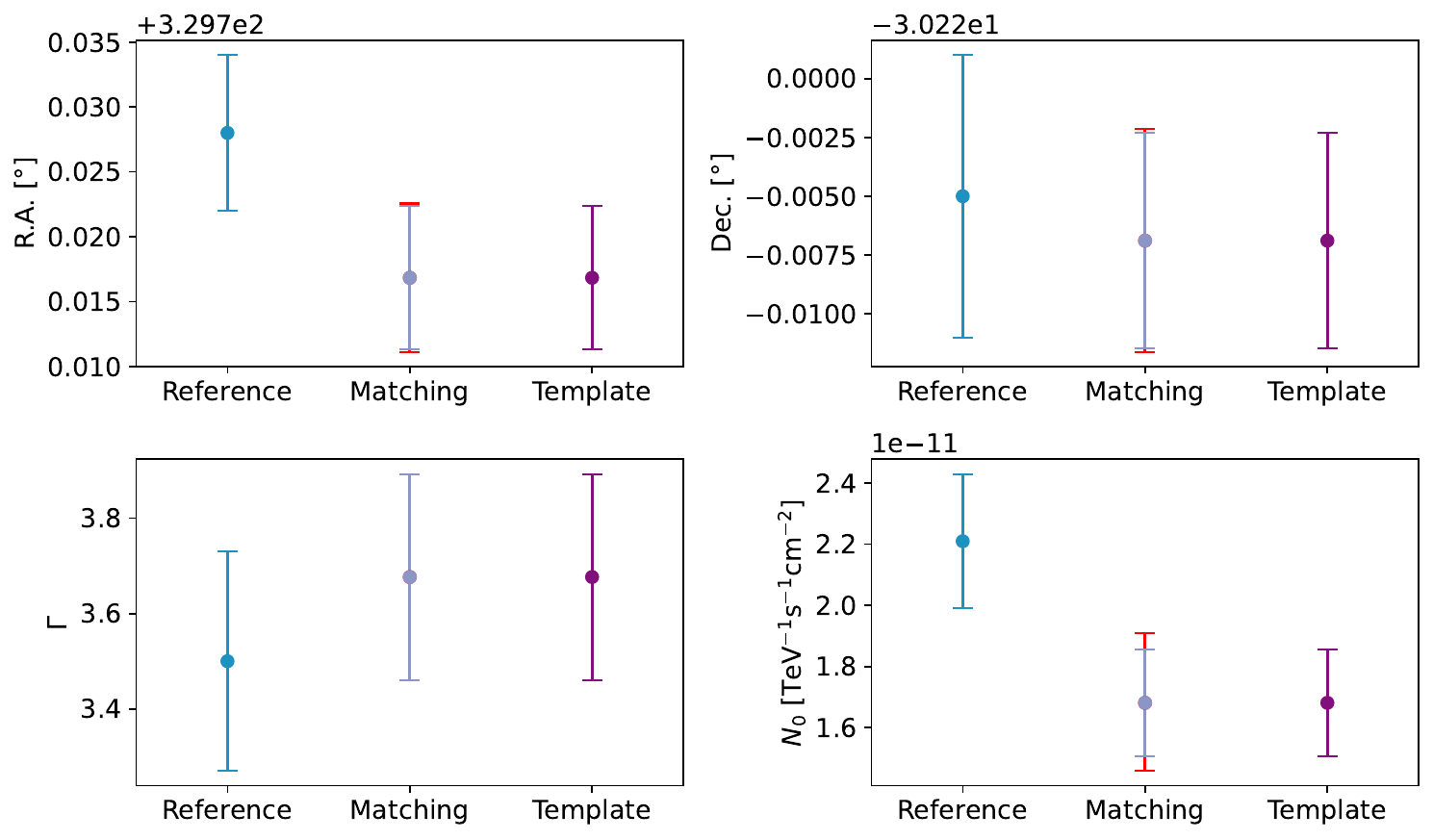}
\caption{Comparison between the best-fit values for all parameters of the point model used to describe the emission from PKS~2155$-$304. The reference values where taken from \citet{bkg_template}, Matching refers to the values derived by using the run-matching approach for the background estimation, while template refers to the background model template. The systematics introduced due to the run-matching approach are indicated in red.}
\label{fig:params-pks}
\end{figure*}

\begin{figure*}
\centering
\includegraphics[width=15cm]{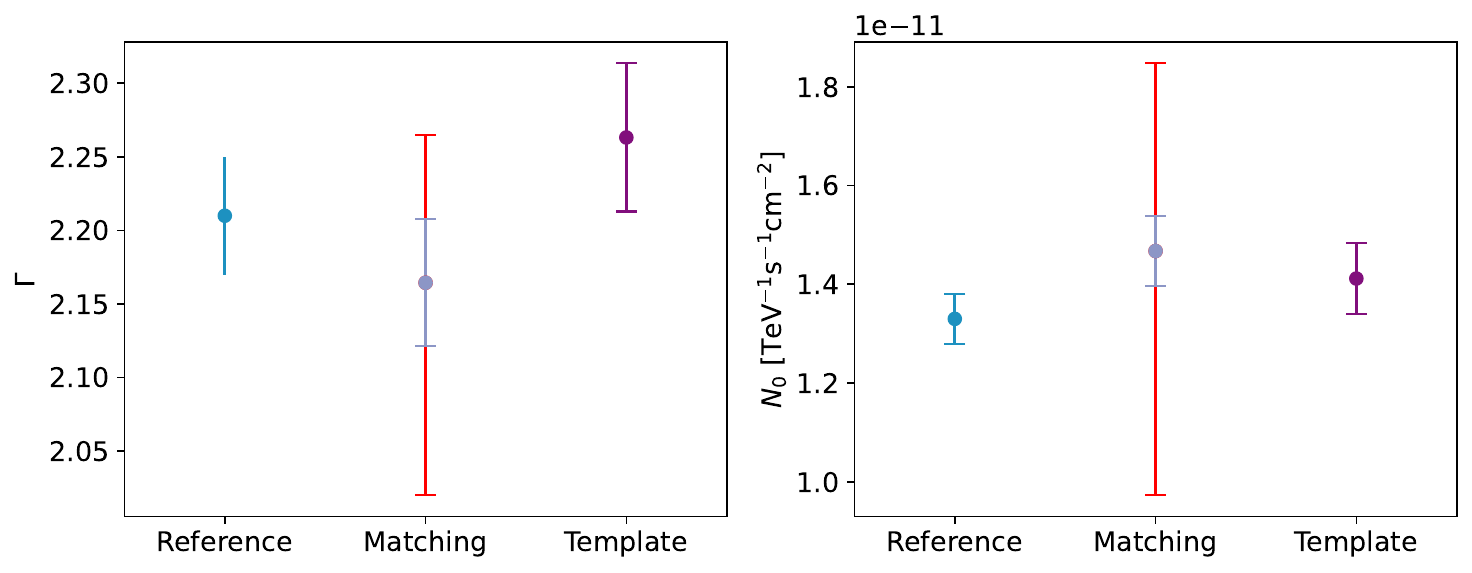}
\caption{Comparison between the best-fit values for all fit parameters of the model used to describe the emission from RX~J1713.7$-$3946. The reference values where taken from \citet{bkg_template}, Matching refers to the values derived by using the run-matching approach for the background estimation, while template refers to the background model template. The systematics introduced due to the run-matching approach are indicated in red.}
\label{fig:params-rxj}
\end{figure*}

Table \ref{tab:fit-params} contains the best-fit values obtained using the background model template and run-matching approaches, for all source regions analysed in this study. Additionally, the best-fit values from \citet{bkg_template} are listed for comparison. Figure \ref{fig:params-crab}, Figure \ref{fig:params-pks} and Figure \ref{fig:params-rxj} show a visual comparison of the best-fit values including the systematic errors for the regions around the Crab Nebula, PKS~2155$-$304 and RX~J1713.7$-$3946 respectively. 

\clearpage
\begin{landscape} 
\begin{table}
\caption{The best-fit parameters for all sources analysed in this study.} 
\label{tab:fit-params} 
\centering    
\begin{threeparttable}
\begin{tabular}{c c | c c c c c c c c} 
\hline\hline  
 &  & RA $[^\circ]$ & Dec $[^\circ]$ & $r_0$ $[^\circ]$ & $e$ $[^\circ]$ & $\phi$ $[^\circ]$ & $\text{E}_0$ [TeV] & $\text{N}_0$ $[10^{-12} \,\text{cm}^{-2}\, \text{s}^{-1}\,\text{TeV}^{-1}]$ & $\Gamma$ \\ 
 \hline
\rule{0pt}{4ex} 
            & Reference & $83.619 \pm 0.003$ & $22.025 \pm 0.002$  & $ -- $ & $--$ & $--$ & $1.45$ & $17.3 \pm 0.8$ & $2.69 \pm 0.07$  \\
Crab Nebula & Case~0    & $83.628 \pm 0.002$ & $22.026 \pm 0.002$ & $--$ & $--$ & $--$ & $1.45$ & $16.8 \pm 0.7$ & $2.73 \pm 0.06$ \\
            & Case~3    & $83.629 \pm 0.002$ & $22.027 \pm 0.002$ & $--$ & $--$ & $--$ & $1.45$ & $16.1 \pm 0.7$ & $2.67 \pm 0.06$ \\

    \\
\hdashline
\\
               & Reference & $329.73 \pm 0.007$  & $-30.227 \pm 0.006$ & $--$ & $--$ & $--$ & $0.65$ & $22.1 \pm 2.2$ & $3.50 \pm 0.23$  \\
PKS~2155$-$304 & Case~0    & $329.72 \pm 0.005$ & $-30.224 \pm 0.004$ & $--$ & $--$ & $--$ & $0.65$ & $17.1 \pm 1.8$ & $3.73 \pm 0.21$ \\
               & Case~3    & $329.72 \pm 0.006$ & $-30.227 \pm 0.005$ & $--$ & $--$ & $--$ & $0.65$ & $16.8 \pm 1.7$ & $3.68 \pm 0.2$ \\
    \\
\hdashline
\\
                     & Reference & $228.55 \pm 0.01$ & $-59.171 \pm 0.007$ & $0.196 \pm 0.007$ & $0.801 \pm 0.038$ & $151 \pm 6.0$ & $1.4$ & $2.91 \pm 0.14$ & $2.41 \pm 0.05$  \\
MSH~15$-$5\textit{2} & Case~0    & $228.56 \pm 0.04$ & $-59.175 \pm 0.003$ & $0.170 \pm 0.004$ & $0.739 \pm 0.020$ & $145.4 \pm 2.2$ & $1.4$ & $3.35 \pm 0.24$ & $2.42 \pm 0.08$ \\
                     & Case~3    & $228.56 \pm 0.02$ & $-59.170 \pm 0.001$ & $0.169 \pm 0.001$ & $0.765 \pm 0.034$ & $149 \pm 7.8$ & $1.4$ & $3.26 \pm 0.34$ & $2.40 \pm 0.08$ \\
    \\
    \hdashline
    \\
                  & Reference & $--$ & $--$ & $--$ & $--$ & $--$ & $1.15$ & $13.3 \pm 0.5$ & $2.21 \pm 0.04$ \\
RX~J1713.7$-$3946 & Case~0    & $--$ & $--$ & $--$ & $--$ & $--$ & $1.15$ & $14.1 \pm 0.7$ & $2.26 \pm 0.05$ \\
                  & Case~3    & $--$ & $--$ & $--$ & $--$ & $--$ & $1.15$ & $14.9 \pm 0.7$ & $2.17 \pm 0.04$ \\
\end{tabular}
\begin{tablenotes}
            \item \textbf{Note:} The parameters $r_0$, $e$, $\Phi$ are the major axis, eccentricity, and position angle of the elliptical disk used to describe the emission from MSH~15$-$5\textit{2}. These fit parameters are compared to reference values from \citet{bkg_template}. The reference energy $E_\text{ref}$ was fixed for the likelihood-minimisation.
        \end{tablenotes}
\end{threeparttable}
\end{table}
\end{landscape}

\section{Additional significance maps and spectra}
\begin{figure}[h]
\centering
\includegraphics[width=0.49\textwidth]{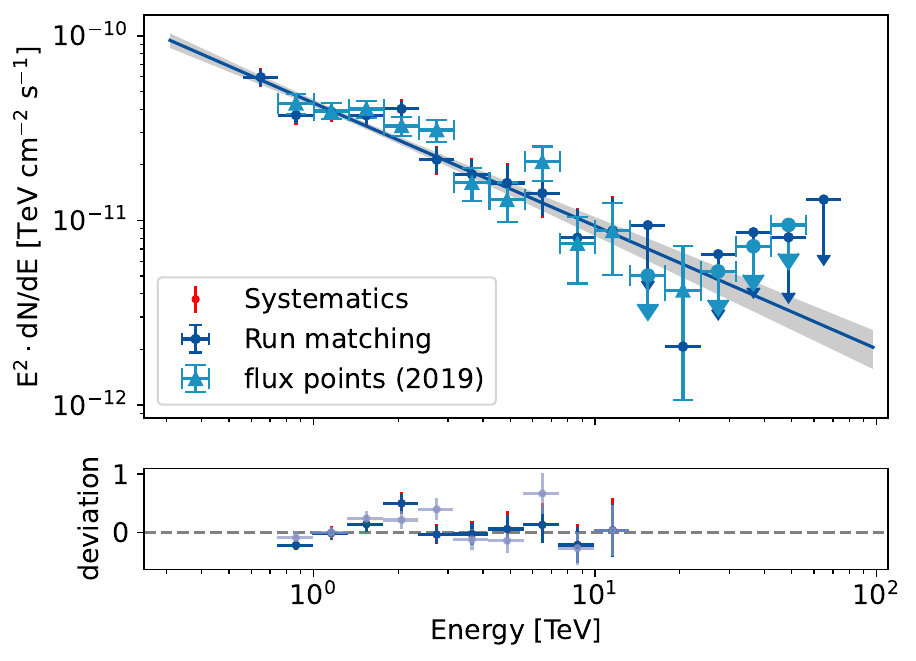}
\caption{SED of the Crab Nebula using the run-matching approach for the background estimation compared to the reference SED from \citep{bkg_template}. The deviation between the two SEDs and the best-fit model derived for the Case~3 datasets is depicted in the lower panel.}
\label{fig:sed_crab}
\end{figure}

\begin{figure}[h]
\centering
\includegraphics[width=0.49\textwidth]{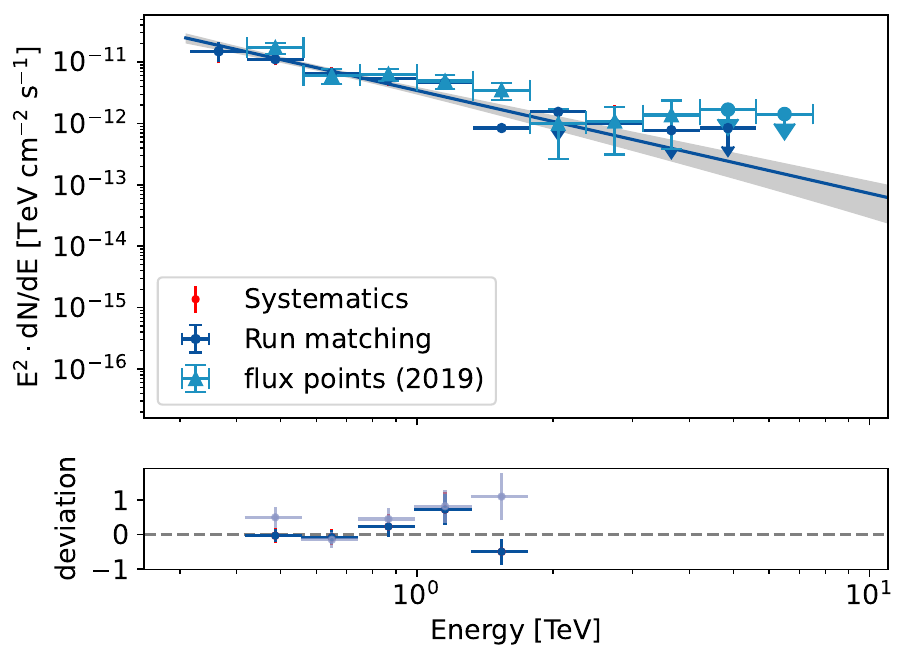}
\caption{As in Fig. \ref{fig:sed_crab} for PKS~2155$-$304}
\label{fig:sed_pks}
\end{figure}

\begin{figure}[h]
\centering
\includegraphics[width=0.49\textwidth]{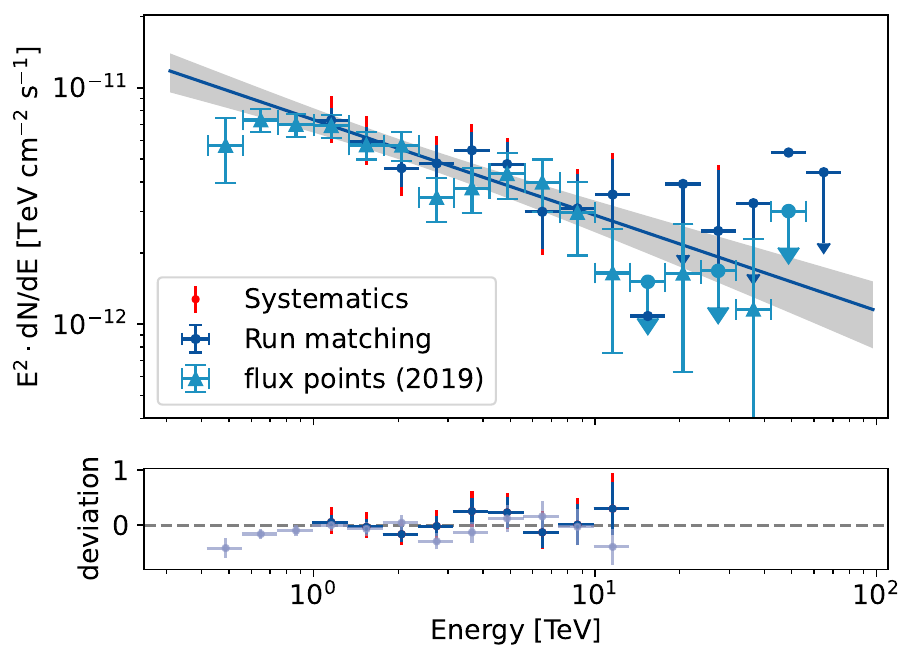}
\caption{As in Fig. \ref{fig:sed_crab} for MSH~15$-$5\textit{2}}
\label{fig:sed_msh}
\end{figure}

\begin{figure*}[h]
\begin{minipage}[b]{.49\textwidth}
\includegraphics[width=0.9\textwidth]{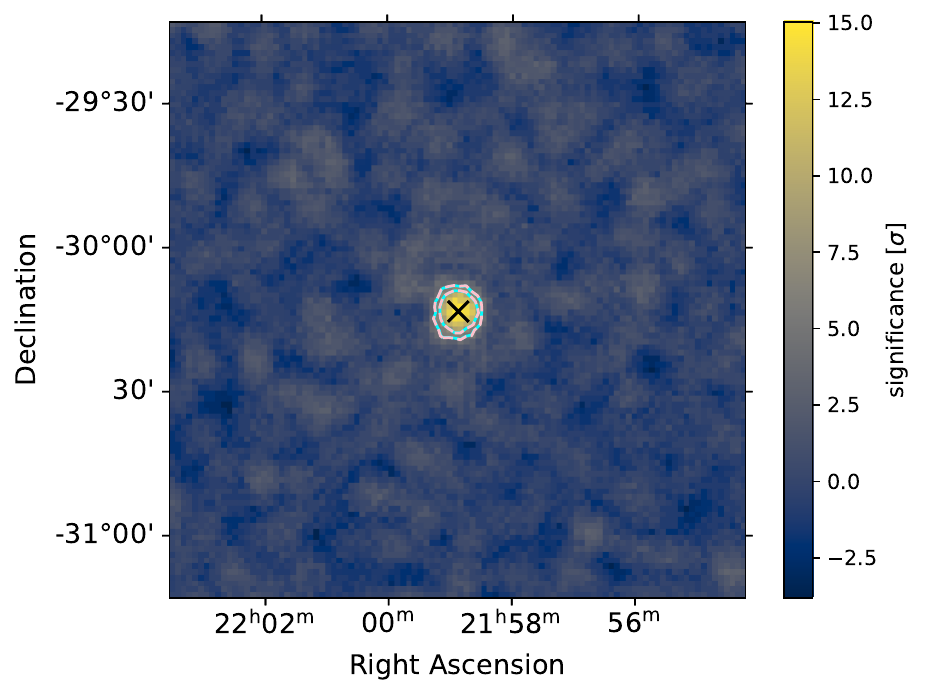}
\end{minipage}\qquad
\begin{minipage}[b]{.49\textwidth}
\includegraphics[width=0.9\textwidth]{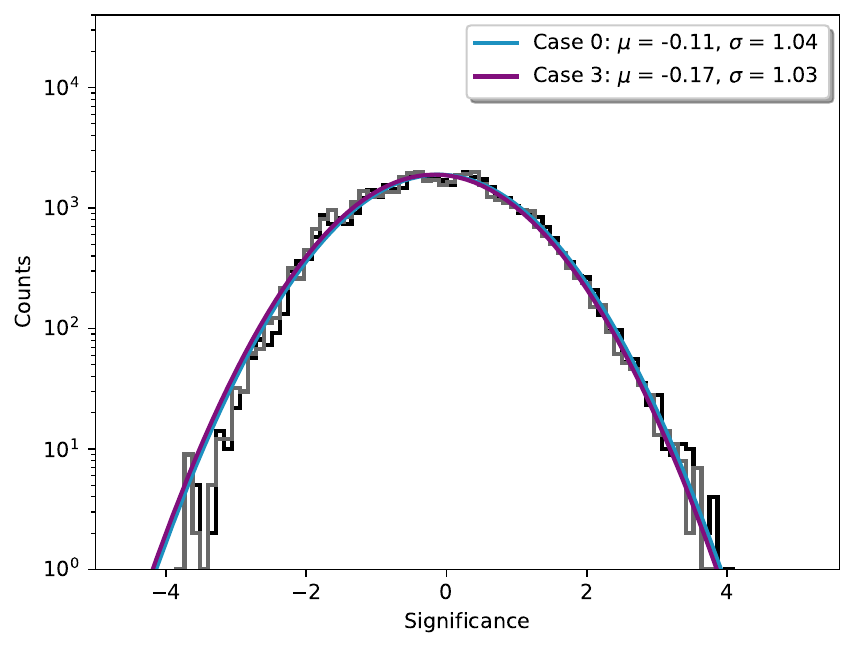}
\end{minipage}
\caption{Left: Li\&Ma significance map of the region around PKS~2155$-$304, with $5\,\sigma$ and $8\,\sigma$ contours,  from the Case~0 dataset (blue), and from the Case~3 dataset (pink). The best-fit position is indicated by the black cross. Right: significance distribution of the background in the FoV around PKS~2155$-$304 for both background estimation techniques, fit with a normal distribution.}
\label{fig:pks_sigmap}%
\end{figure*}

\begin{figure*}[h]
\begin{minipage}[b]{.49\textwidth}
\includegraphics[width=0.9\textwidth]{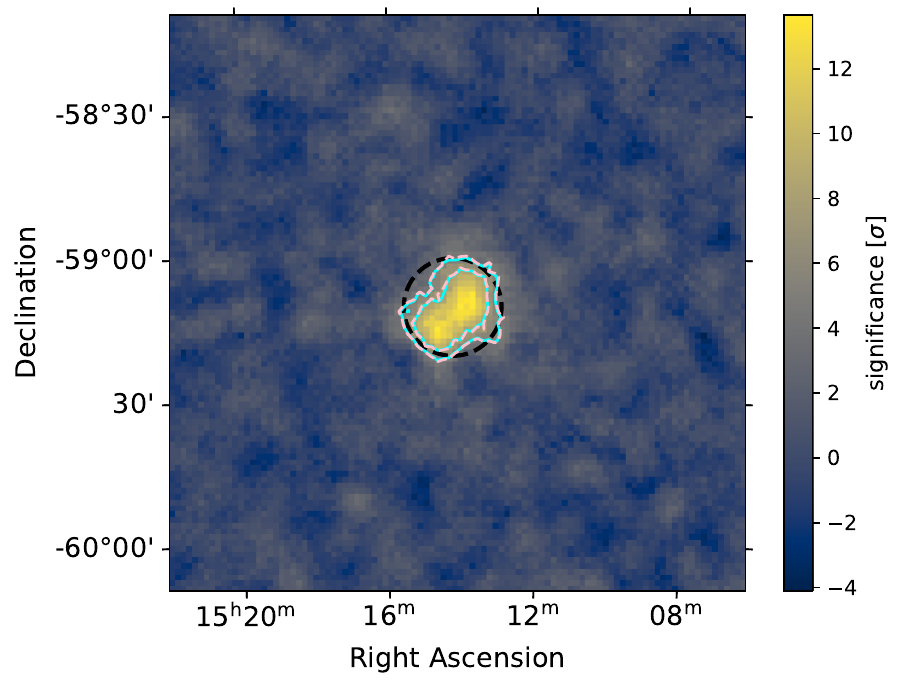}
\end{minipage}\qquad
\begin{minipage}[b]{.49\textwidth}
\includegraphics[width=0.9\textwidth]{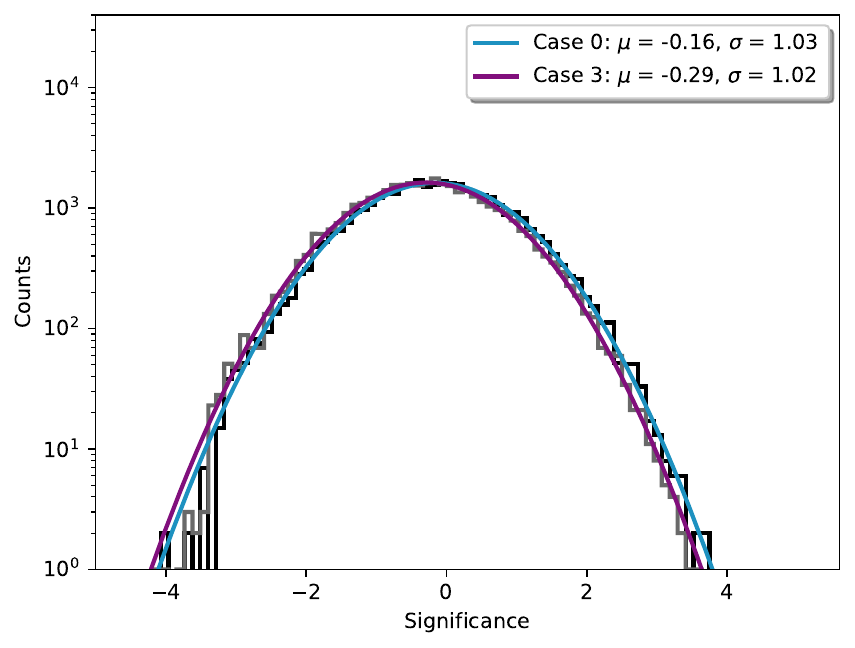}
\end{minipage}
\caption{Left: Li\&Ma significance map of the region around MSH~15$-$5\textit{2}, with $5\,\sigma$ and $8\,\sigma$ contours from the Case~0 dataset (blue) and from the Case~3 dataset (pink). The best-fit morphology is indicated by the black-dashed line. Right: The significance distribution of the background in the FoV for both background estimation techniques, fit with a normal distribution.}
\label{fig:msh_sigmap}%
\end{figure*}

\begin{figure*}[h]
\begin{minipage}[b]{.49\textwidth}
\includegraphics[width=0.9\textwidth]{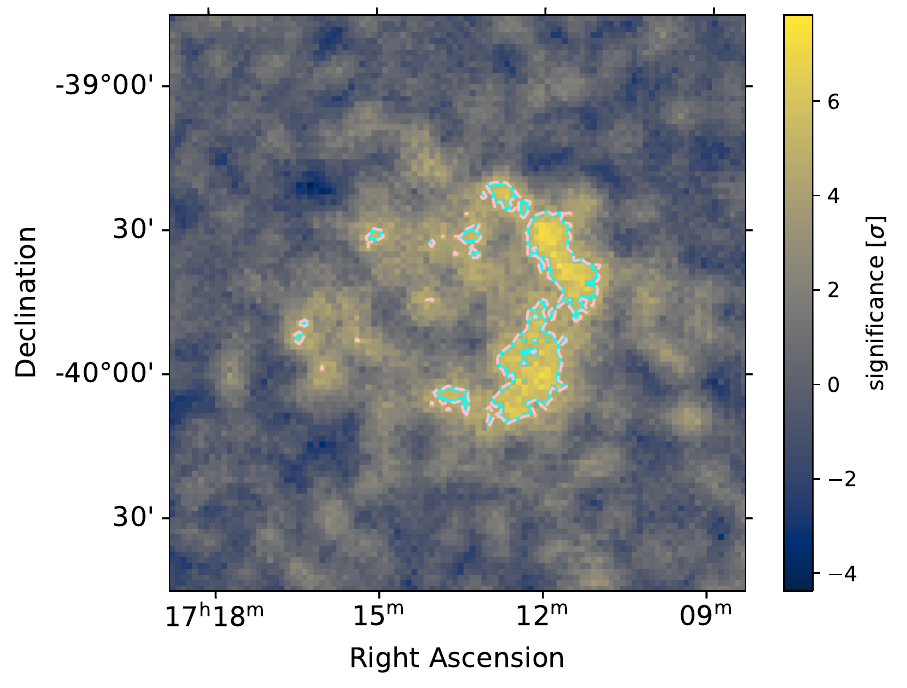}
\end{minipage}\qquad
\begin{minipage}[b]{.49\textwidth}
\includegraphics[width=0.9\textwidth]{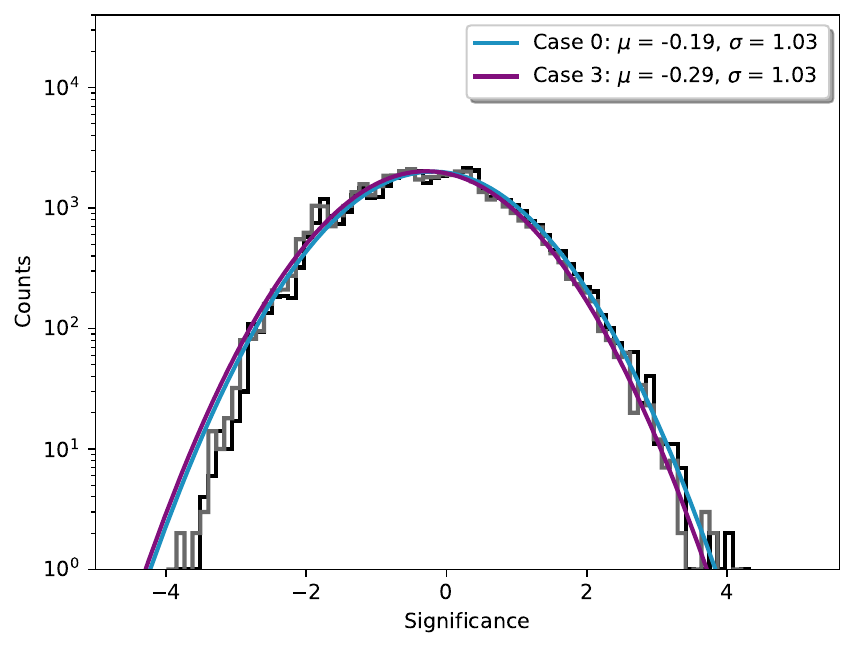}
\end{minipage}
\caption{Left: Li\&Ma significance map of the region around RX~J1713.7$-$3946, with $5\,\sigma$ and $8\,\sigma$ contours for Case~0 (blue), and for Case~3 (pink). Right: significance distribution of the background in the FoV for both background estimation techniques, fit with a normal distribution.}
\label{fig:rxj_sigmap}%
\end{figure*}

Here we show the significance maps and distributions of the region around PKS~2155$-$304 (Figure \ref{fig:pks_sigmap}), MSH~15$-$5\textit{2} (Figure \ref{fig:msh_sigmap}) and RX~J1713.7$-$3946 (Figure \ref{fig:rxj_sigmap}). Although a small shift in the background distributions for the datasets estimated using the run-matching approach can be observed, this shift is within the systematic uncertainty (see Table \ref{tab:background_hist}).
\\
The SED estimated for the Crab Nebula, PKS~2155$-$304 and MSH~15$-$5\textit{2} are shown in Figures \ref{fig:sed_crab}, \ref{fig:sed_pks} and \ref{fig:sed_msh} respectively. The lower panel of these figures shows the deviation between the two sets of fluxpoints and the best-fit spectral model derived in the analysis of the respective Case~3 datasets, defined by $(x_{1/2} - x_{model})/x_{model}$, with $x_{1}$ the differential energy flux in the respective energy bin for the reference fluxpoints derived in \citet{bkg_template}, $x_{2}$ the differential energy flux for the fluxpoints derived from the Case~3 datasets and $x_{model}$ the differential energy flux estimated from the best-fit spectral model.

\end{appendix}

\end{document}